\documentclass[]{article}
\usepackage[a4paper, total={5.5in, 8in}]{geometry}
\usepackage{graphicx}
\usepackage{color,soul}
\usepackage{cancel}
\usepackage{amsmath}
\usepackage{amsfonts}
\usepackage[mathlines]{lineno}
\usepackage{courier}
\usepackage{multirow}

\newcommand{\interior}[1]{%
	{\kern0pt#1}^{\mathrm{o}}%
}

\definecolor{giu_comm}{rgb}{0.0, 0.42, 0.24}
\DeclareUnicodeCharacter{0308}{HERE!HERE!}

\newcommand{\BibTeX}{{\rm B\kern-.05em{\sc i\kern-.025em b}\kern-.08em\ignorespaces
  T\kern-.1667em\lower.7ex\hbox{E}\kern-.125emX}}

\begin{document}
\overfullrule=9pt

\title{\Large{\textbf{Hypoxia-related radiotherapy resistance in tumours: treatment efficacy investigation in an eco-evolutionary perspective.}}}%
\author{Giulia Chiari$^{*,1,2,3}$, Giada Fiandaca$^{4}$, Marcello Delitala$^{1}$}

\maketitle
\begin{center}
$^{1}$ Department of Mathematical Sciences ``G. L. Lagrange'', Politecnico di Torino (Turin, Italy)\\
$^{2}$ Department of Mathematics ``G. Peano'', Università di Torino (Turin, Italy)\\
$^{3}$ Department of Mathematics, Swinburne University of Technology (Melbourne, Australia)\\
$^{4}$ Department of Cellular, Computational and Integrative Biology, Università degli Studi di Trento (Trento, Italy)\\    
\end{center}
%
%

\begin{abstract}
In the study of therapeutic strategies for the treatment of cancer, eco-evolutionary dynamics are of particular interest, since characteristics of the tumour population, interaction with the environment and effects of the treatment, influence the geometric and epigenetic characterization of the tumour with direct consequences on the efficacy of the therapy and possible relapses. In particular, when considering radiotherapy, oxygen concentration plays a central role both in determining the effectiveness of the treatment and the selective pressure due to hypoxia.\\
We propose a mathematical model, settled in the framework of epigenetically- structured population dynamics and formulated in terms of systems of coupled non-linear integro-differential equations, that aims to catch these phenomena and to provide a predictive tool for the tumour mass evolution and therapeutic effects.\\
The outcomes of the simulations show how the model is able to explain the impact of environmental selection and therapies on the evolution of the mass, motivating observed dynamics such as relapses and therapeutic failures. Furthermore it offers a first hint for the development of therapies which 
can be adapted to overcome problems of resistance and relapses.
\end{abstract}
\vspace{0.5cm}
\textit{keywords:} radiotherapy resistance,  hypoxia, tumour phenotypic heterogeneity, tumour-microenvironment interaction, niche construction, continuous structured models, tailored radiotherapy protocols.\\
\rule{\textwidth}{0.4pt}
\begin{center}
    correspondent author {\texttt{giulia.chiari@polito.it}}
\end{center}

\newpage

\section{Introduction}
Since the recognition of the malignancy, the objective of cancer research is to discover novel methods of quality treatment approaches for eradicate it. Presently, due to its high diffusion, over 60$\%$ of all ongoing medical quality treatment trials worldwide are concentrating on it, \cite{abbas2018overview}. 
\indent Because of its high cytotoxic potential, radiation therapy is a standard of care in many solid tumours, \cite{baskar2012cancer}. 
Its main goal is to deprive cancer cells of their multiplication potential damaging their genetic material and thus blocking their ability to divide and proliferate further \textit{via} high-energy radiation. Although both cancer and normal cells bear its action, tumour cells are in general not as efficient as normal cells in repairing the damage caused by radiation resulting in differential cell killing. For this reason, radiotherapy is mainly delivered through fractionated schemes to maximize the radiation effects to abnormal cancer cells while minimizing exposure to normal ones, \cite{baskar2012diverse}. 
\indent Prediction of tumour response after irradiation has been a challenge at the very beginning since it became rapidly clear that the biological effect of irradiation is a complex phenomenon non uniquely determined by the total dose, but also by the characteristics of the treatment protocol (such as fraction dose and dose schedule) as well as by physiological conditions in which it is applied that can widely range between patients, \cite{demaria2012radiation}.
\indent The success of radiotherapy depends indeed on multiple sub-cellular, cellular, and microenvironmental parameters, together referred to as the “\textit{6Rs
of radiation therapy}”: \textit{repair} of irradiation-induced DNA damages, \textit{redistribution} of cells within the cell cycle, \textit{repopulation} of mass after radiation, \textit{reoxygenation} of the tumour microenvironment, intrinsic \textit{radiosensitivity} of different cell subpopulations and \textit{reactivation} of the anti-tumour immune response, \cite{rakotomalala2021hypoxia}.\\
\indent A crucial factor that impacts on all these aspects is tumour heterogeneity in terms of both microenvironmental conditions and cancer cell populations. In particular, it has been observed that the local oxygen concentration can significantly influence radiation induced cell death, with well-oxygenated regions being shown to exhibit up to threefold greater radiosensitivity than hypoxic tumour populations, \cite{horsman2012imaging}.\\ 
\indent Hypoxia is a consequence of the high tumour cell proliferation rate and the abnormal structure of the tumour vasculature. Oxygenation level is generally reduced and heterogeneous within malignant masses, compared to the oxygenation found in associated healthy tissues; this oxygen lack is a critical features in tumours promoting their progression. It is indeed clinically observed that, in solid tumours, oxygen tissue deprivation acts as an environmental stressor, promoting a long series of genetic, but especially, epigenetic mutations that strongly impact the tumour eco-evolutionary dynamics. Cancer cells are indeed able to adjust their cellular physiology and metabolism via the up-regulation of different genes as p53, HIF-$\alpha$ or GLUT-1 or IAP-2 acquiring the ability to grow in hypoxic microenvironments and to evade apoptosis, \cite{fukumura2007tumor}. \\
\indent The reason for which low oxygen tensions are associated with radio-resistance relies on the mechanism of cell killing by ionizing radiation. It is indeed experimentally shown that oxygen plays a fundamental role in fixing the damage on cancer cells induced by radiotherapy that lead to their death, \cite{kim2019cellular}.\\ 
In fact, it is observed that hypoxia can cause topographically defined cellular subpopulations protected at the time of radiation without being killed by severe oxygen starvation; the oxygen tension for hypoxic cells could be indeed high enough to allow for clonogenic survival, but low enough to protect them from the effects of ionizing radiation, \cite{rockwell2009hypoxia}.\\
\indent In this view, it is clear that hypoxia impacts all the \textit{6Rs} mentioned before, becoming a fundamental factor to be consider to a successful treatment protocol. In particular, we are interested in investigating its effects on three of them: (\textit{i}) \textit{radiosensitivity}, (\textit{ii}) \textit{repopulation} and (\textit{iii}) \textit{reoxygenation}. \textit{Radiosensitivity} defines the intrinsic sensitivity of tumour cells to the therapy; it is influenced by hypoxia at two levels: a \textit{direct} one since, as underlined before, oxygen is responsible for the enhancement of the detrimental effect of ionizing radiation which implies that radiotherapy is less efficient in the areas in which a lack of oxygen is observed; an \textit{inverse} one by the fact that hypoxia selects for cells equipped by high resistance to hostile environments and low proliferative rates. 
\textit{Repopulation} defines instead the renewal and proliferation of surviving cancer cells following irradiation and is affected by hypoxia since it promotes treatment resistant hypoxic cells that serve as a nidus for subsequent tumour regrowth and repopulation. Finally, \textit{reoxygenation} defines the fact that, between radiotherapy fractions, well-oxygenated cells death leads to oxygen release, reduction of oxygen demand, and tumour bulk shrinkage allowing better oxygen diffusion turning back initially
refractory hypoxic areas to a more radiosensitive state, \cite{rakotomalala2021hypoxia}.\\
\indent The emergence of a resistant population can be described in terms of tumour evolution and stems from its intrinsic heterogeneity. In an eco-evolutionary perspective, a tumour can be indeed interpreted as a cells population characterized by an accumulation, via natural selection, of genetics and epigenetics alterations that appear both due to intrinsic cells variability and to their mutual interactions with the surrounding microenvironment. In this light, all the treatment procedures could act as a environmental stressor on tumour cells inducing strong modifications of tumour ecology and, consequently, of the fitness landscape of tumour cells promoting variations tumours composition. The resulting strong selective bottleneck enriches for resistant phenotypes within cancer cells as mirror of the evolutionary capacity of cancer phenotypes to adapt to therapeutic perturbations as well as of the modifications of the temporal and spatial heterogeneity of the tumour microenvironment (\cite{gatenby2020integrating,burrell2014tumour,dujon2021identifying}). 
In this view, ecologically informed therapeutic strategies can potentially define and use novel treatment approaches that could vary among patients whose landscapes could be completely different. Such an adaptive approach implies that each patient therapeutic protocol is strictly personalized on the basis of the tumour state and response rather than a one-size-fits-all fixed treatment regime, (\cite{brady2021predicting, corwin2013toward}).\\
\indent Mathematical models constitute a good investigation instrument in this sense
since they can allow to test different environmental conditions, different tumour
compositions as well as different therapeutic protocols. They can be seen as \textit{in silico} laboratories to evaluate the mutual interactions between the above mentioned aspects and their consequences on tumour development. 
In this respect, the effect of tumour-host interaction, in particular considering tissue oxygenation and its role on shaping the phenotypic composition of tumour masses and their double impact on radiotherapy efficacy has been deeply investigated \textit{via} a wide range of modeling techniques, \cite{ardavseva2020mathematical,strobl2020mix,villa2021evolutionary,lorenzi2018role,lorz2015modeling,villa2021modeling, fiandaca2020mathematical}. As example, in \cite{celora2023spatio}, the authors presented a mathematical model that describes how tumour heterogeneity in terms of stemness evolves in a tissue slice oxygenated by a single blood vessel, determining the proliferative capacity, the apoptosis propensity and the response to radiotherapy protocols. A similar dynamics is investigated in \cite{hamis2019blackboard} via a hybrid cellular automaton in which the authors analysed the spatio-temporal dynamics and the evolution of the intratumoural heterogeneity of a mass under the action of radiotherapy, showing how the treatment results more effective in well-oxygenated tumours than in 
the poorly oxygenated ones. In the same veins but with a particular attention to tumour cell repopulation, reoxygenation and redistribution of proliferative states, in \cite{kuznetsov2020optimization}, it is proposed a spatially-distributed continuous matematical model of solid tumour growth treated by fractionated RT. 
Other interesting results are collected in \cite{lewin2018evolution} in which numerical and analytical techniques are developed to investigate radiation response of tumours with different intratumoural oxygen distribution profile. Finally, without an explicit description of tumour oxygenation but in a more general framework of tumour-host interaction in terms of competition for space and resources and tumour heterogeneity, in \cite{poleszczuk2018predicting} the authors proposed a prognostic factor for personalized radiotherapy, named Proliferation Saturation Index (PSI), to identify the best fractionation scheme.\\
Following this research line, a particularly promising treatment modality is the Intensity-Modulated RadioTherapy (IMRT), which has the potential to be an effective method for delivering customized radiation therapy to small, specific regions of a tumour based on its oxygenation level, \cite{ling2000towards,baumann2016radiation}. This approach is called \textit{dose painting} and involves selectively boosting doses to regions of the tumour that are known to be particularly resistant to treatment, \cite{grimes2017hypoxia}. To fully exploit this technique, however, additional information about the tumour composition, specifically in terms of its resistance to hypoxia, is necessary in order to define the most effective radiation dosimetry plan.\\
\indent Motivated from the above considerations, in this work, we are interested in investigating how low oxygen levels and hypoxia-associated tumour cell adaptions affect radiotherapy efficiency in the specific case of solid tumours. We aim, in this sense, to develop a tool which could adapt to patient specific characteristics, in line with the innovative personalized medicine approach (\cite{flashner2019decoding}).\\

\indent In our previous work presented in \cite{chiarihypoxia}, we deeply investigated how the mutual interactions between the tumour mass and oxygen distribution, (\textit{i}) can result in a geometric characterization of tumour niches in terms of masses spatial extension, how this characterization could affect the phenotypic composition in terms of survival and invasive abilities and finally how both these two aspects in synergy affect the mass growth. This approach naturally laid the groundwork to investigate how the pre-therapeutic history of a tumour dictated by oxygen distribution could determine therapeutic failures thanks to the possibility to take into account the differences in tumour conformation and invasive ability coupled with the emergence of treatment resistant hypoxic cells that result from this dynamics. It indeed perfectly matches with the necessity to take into account two crucial events: (\textit{i}) hypoxia selects for cells equipped by high resistance to hostile environments and low proliferative rates; (\textit{ii}) these cells are intrinsically less exposed to treatment action with respect the normal cells being the power of radiotherapy to be able to block the replication process.  
\indent This setting clearly results to be particularly suitable to investigate \textit{radiosensitivity} development dictated by hypoxia since we can naturally map cell mitotic potential with their intrinsic resistance ability. A \textit{low proliferant} cell is indeed intrinsically \textit{more resistant} to the action of radiotherapy in light of what we previously observed. Moreover, the eco-evolutionary approach that we there adopted allows also to investigate the effects on tumour growth and regrowth of therapeutic perturbations coupled with the spatial and temporal variations observed in tumour microenvironment leading to investigate also the dynamics that governs the \textit{repopulation} of a tumour mass after the treatment. Finally, considering the interaction between the tumour mass and the microenvironment in which it lives, it allows also to focus on the dynamics of oxygen and to evaluate the impact of \textit{reoxygenation} phenomena.\\ 
\indent In this work, we present an extension of the model in \cite{chiarihypoxia} to take into account how the effects of radiotherapy differ according to the heterogeneity faced at the instant and the location at which the therapy is applied (from both a physical and a phenotypical point of view) investigating how a divergent response could be observed within and among patients. In this perspective, we consider a specific formulation for the survival fraction of the already treated tumour cells, able to capture both parameters directly associated with clinical data and specific mortality rates with respect to different doses and treatment timings. This new modeling structure of radiotherapy, gives the possibility to explore the tumour-therapy interaction in two mutual directions i.e. (\textit{i}) the impact of tumour developmental dynamics on the efficacy of therapy, (\textit{ii}) the impact of therapy on the spatial and epigenetic characteristics of the tumour mass.\\
Moreover, the action of the environmental selection is taken into account to characterize the spatial heterogeneity of proliferative potential and to identify the tumour regions composed by cell with low proliferative rate and to study how their evolution could strongly influence treatment success.\\
Since the terms phenotypic and epigenetic (both already introduced in this work) are often used in the literature with the same meaning, we specify that in the sequel of this paper, we refer to epigenetic trait (and the relative mutation) when we intend to refer to the molecular imprint on the genotype which determines the degree of activation of the genes, keeping unaltered their sequence. Instead, we refer to the phenotype as the observable actualization of interactions between its genome, epigenome, and local environment.\\
\indent The rest of the paper is organized at it follows: in Section \ref{sec:Materials and methods}, we present the proposed model with the underlying assumptions (see Subsection \ref{subsec:Mathematical model}); details on the parameters estimate and on its numerical implementation as well as on the indices that quantity tumour progression are given in Subsection \ref{subsec:Simulation details} and Subsection \ref{subsec:Quantification of model results}, respectively. We then turn on describing the model results in Section \ref{sec:Results}. Specifically, we simulate the growth of the malignancy in two specific settings, referred as Case 1 -  {highly oxygenated tissue} (see Subsection \ref{subsec:Case 1 - highly efficient single vessel}) and Case 2 -  {poorly oxygenated tissue} (see Subsection \ref{subsec:Case 2 - inefficient single vessel}) that differ with respect to the oxygenation level of the tissue. We compare them applying the same radiotherapy protocol to highlight the differences that could be observed in tumour response due to tumour-host interaction. Subsection \ref{subsec:Dose painting} is instead devoted to investigate possible variations of radiotherapy efficacy varying the total dose amount delivered in the two experimental setting toward tailored radiotherapy protocols. 
Finally, in Subsection \ref{subsec:Heterogeneous vasculature}, we analyse the effect of spatially heterogeneous distributions of the intra-tumoural  {oxygen sources} to highlight their role in the creation of ecological niches, due to the relative blood vessels dispersal that influence the treatment response. The article ends in Section \ref{sec:Conclusions} with a summary of the main results with hints for possible developments.
\section{Materials and methods}
\label{sec:Materials and methods}

\subsection{Mathematical model}
\label{subsec:Mathematical model}
As mentioned in the introduction, building upon our previous work \cite{chiarihypoxia}, we extend the model to include the effects of radiotherapy. To this aim, we set a spatial bi-dimensional domain $\Omega_s \subset \mathbb{R}^2$ in which the mass can expand assuming to observe a tumour evolving in a tissue slice. In particular, in our setting: 
(\textit{i}) oxygen is the main environmental actor to affect tumour evolution, and in the determination of the different areas of therapeutic efficacy; (\textit{ii}) tumour cells behaviour will be influenced by the epigenetic characteristics of individuals in terms of their double \textit{resistance} to hypoxia and radiotherapy, the environmental conditions faced and the mutual interaction between these two aspects.  
\indent In this respect, the virtual tumour mass is differentiated in \textit{metabolically active} (\textit{i.e.}, viable) and \textit{necrotic} individuals. The necrotic subpopulation is assumed to be undifferentiated, with number density given by the function $n(t,\mathbf{x}): T \times \Omega_s \mapsto \mathbb{R}^{+}_{0}$. The viable tumour portion is instead structured with respect to the epigenetic trait $u \in \Omega_p= [0,1]$ that describes the \textit{double} resistance level of malignant cells \textit{i.e.} w.r.t. both the ability to \textit{survive in hypoxic tumour areas} and their \textit{radiosensitivity}. A choice of this type allows to take into account a continuum spectrum of possible cell states characterized by intermediate levels of both survival and proliferation in such a way that an increasing epigenetic expression denotes an higher survival ability and a lower duplication rate. In particular, the epigenetic state $u=0$ will characterize the cells that show the highest mitotic potential and, relatively, the highest sensibility to both oxygen lack and radiotherapy action (\textit{proliferation promoting or sensible cells}); the epigenetic state $u=1$, instead is linked to cells that show the potentially highest survival ability but the lowest duplication capacity (\textit{survival promoting or resistant cells}).\\
\indent The tumour mass distribution on the spatial and epigenetic space is identified with the function $a(t,\mathbf{x},u): T \times \Omega_s \times \Omega_p \mapsto \mathbb{R}^{+}_{0}$ that reflects the epigenetic composition of the tumour mass located at time $t$ in the domain point $\mathbf{x}$. Its evolution is modelled \textit{via} the following trait-structured integro-differential equation (IDE):
\begin{equation}
	\label{eq:a}
	\frac{\partial a(t,\mathbf{x},u)}{\partial t} = \underbrace{\mu_{p} \, \frac{\partial^2 a(t,\mathbf{x},u)}{\partial u^2}}_{\substack{\textup{epigenetic variations}}} + \underbrace{\mu_s \Delta_\mathbf{x} a(t, \mathbf{x}, u)}_{\textup{movement}}+ \underbrace{R(u,O(t,\mathbf{x}),\rho(t,\mathbf{x}),n(t,\mathbf{x}))a(t,\mathbf{x},u)}_{\textup{proliferation/selection/necrosis}}.
\end{equation}
in which $O(t,\mathbf{x}): T \times \Omega_s \mapsto \mathbb{R}^{+}_{0}$ represents the oxygen concentration and $\rho(t,x)$ denotes the local number density of the non-necrotic tumour area, computed considering all the individuals present in the mass regardless of their epigenetic trait:
\begin{equation}\label{eq:rho}
	\rho(t,\mathbf{x})=\int_{\Omega_p} a(t,\mathbf{x},u)\, du.
\end{equation}
In this respect, metabolically active cells (\textit{i}) undergo to random phenotypic transitions, (\textit{ii}) randomly move, (\textit{iii}) either proliferate, being subjected to selective pressures by environmental conditions or acquire an irreversible necrotic fate due to both \textit{oxygen lack} or \textit{radiotherapy action}. 
The reaction term in Eq.\eqref{eq:a} expresses local variations in the mass of viable cells due to proliferation, the action of the natural selection and necrosis phenomena due to \textit{oxygen lack} or \textit{radiotherapy action}. The rates at which these phenomena take place are given by the functions $P$, $S$, $N$ and $T$ respectively:
\begin{align}\label{eq:r}
	R(u,O(t, \mathbf{x}), \rho(t,\mathbf{x}),n(t,\mathbf{x}))=&\underbrace{P(u,O(t,\mathbf{x}),\rho(t,\mathbf{x}),n(t,\mathbf{x}))}_{\textup{proliferation}}-\underbrace{S(u,O(t,\mathbf{x}))}_{\textup{selection}} \nonumber\\
 &-\underbrace{N(O(t,\mathbf{x}))}_{\textup{oxygen lack}}-\underbrace{T(u,O(t,\mathbf{x}))}_{\textup{radiotherapy}}.
\end{align}
In details, the proliferation rate $p$ is assumed to depend on (\textit{i}) the individual actual epigenetic trait, (\textit{ii}) the resources availability and (\textit{iii}) the physical limitations determined by the available space. In this respect, we factorize $p$ as follows:
\begin{equation}
	P(u,O(t,\mathbf{x}),\rho(t,\mathbf{x}),n(t,\mathbf{x}))=p_1(u) \,p_2(O(t,\mathbf{x})) \, p_3(\rho(t,\mathbf{x}),n(t,\mathbf{x})).
	\label{eq:p}
\end{equation}
where $p_1$ represents the duplication law and takes into account the variability in cell proliferation w.r.t. their epigenetic trait in the light of the \textit{proliferation-survival trade-off}. 
Individuals characterized by epigenetic state $u=0$ correspond to the cell variant with the highest proliferation rate, denoted as $\gamma_{\textup{max}}$; whereas individuals characterized by epigenetic state $u=1$ correspond to the cell clone which poorly undergo to mitotic events, as quantified by the lowest proliferation rate $\gamma_{\textup{min}}$. 
We assume a linear trend between the epigenetic firm and the mitotic potential:
\begin{equation}\label{eq:p1}
	p_1(u)= (\gamma_{\textup{min}}-\gamma_{\textup{max}})u+\gamma_{\textup{max}}. 
\end{equation}
The term $p_2$ represents the relation between cell duplication rate and available chemical i.e. cells ability to exploit resources. 
Its expression is modelled considering a classical Michaelis-Menten law:
\begin{equation}\label{eq:p2}
	p_2(O(t,\mathbf{x}))=\frac{O(t,\mathbf{x})}{\alpha_{\textup{O}}+O(t,\mathbf{x})},
\end{equation}
Finally, the term $p_3$ models the growth inhibition due to over-crowding 
by means of a logistic-like law:
\begin{equation}\label{eq:p3}
	p_3(\rho(t,\mathbf{x}),n(t,\mathbf{x}))= 1- \frac{\rho(t,\mathbf{x})+n(t,\mathbf{x})}{k},
\end{equation}
where $k>0$ is the local tumour tissue carrying capacity considering both viable and necrotic individuals in space-competition.\\
\indent The term $S(u,O(t,\mathbf{x}))$ in Eq.\eqref{eq:p1} represents instead the death rate induced by oxygen-driven natural selection. 
In details, it models the experimentally observed trade-off between maximizing cell survival in oxygen deprived tissue and maximizing cell mitotic potential, \cite{martinez2012hypoxic}. 
The chosen \textit{proliferation-survival trade-off} results as follows: 
a lower level of gene expression correlates with a lower resistance to hypoxia, and thus a higher death rate; a higher level of gene expression correlates with a larger fitness cost, and thus a lower proliferation rate. In this respect, we virtually identify different regions in which the spatial domain could be divided with respect to the level of oxygenation observed based on the experimental data presented in \cite{korolev2014turning} and \cite{vaupel2001treatment}. Naming $O_M,O_{m}$ and $O_n$ the oxygen level that bound from below \textit{normoxic}, \textit{hypoxic} and \textit{necrotic} areas, we assume that: (\textit{i}) in normoxic environments (i.e. when $O \geq O_M$), the fittest level of gene expression is the minimal one (i.e. $u=0$); (\textit{ii}) in hypoxic environments (i.e. when $O \leq O_{m}$) the fittest level of gene expression is the maximal one (i.e. $u=1$); (\textit{iii}) in moderately-oxygenated environments (i.e. when $O_{m} < O < O_M$), the fittest level of gene expression is a monotonically decreasing function of the oxygen concentration (i.e. it decreases from $u=1$ to $u=0$ when the oxygen concentration increases); (\textit{iv}) in necrotic areas (i.e. when $O \leq O_{n}$) the oxygen availability is insufficient to guarantee cell survival so cells undergo to apoptosis. Under these assumptions and coherently with our previous work \cite{chiarihypoxia}, 
we define the death rate induced by oxygen-driven selection $S(u,O(t,\mathbf{x}))$ as:
\begin{equation}
	S(u,O(t,\mathbf{x}))= \eta_O \, \big(u-\varphi_O(O(t,\mathbf{x}))\big)^2,
	\label{eq:selection_gg}
\end{equation}	
where (\textit{i}) the parameter $\eta_O>0$ is named the \textit{selection gradient} and quantifies the intensity of oxygen-driven selection; (\textit{ii}) the function $\varphi_O(O(t,\mathbf{x}))$ represents the fittest level of expression of the hypoxia/therapy-resistant gene under specific oxygen concentrations, 
where the \textit{fittest epigenetic expression} with respect to the oxygen concentration is:
\begin{equation}
	\varphi_O(O(t,\mathbf{x}))=
	\begin{cases}
		0, \qquad \qquad \qquad \quad \quad O(t,\mathbf{x}) \geq O_M,
		\\
		\displaystyle
		\frac{O_M-O(t,\mathbf{x})}{O_M-O_{m}}, \quad \quad O_{m} < O(t,\mathbf{x}) < O_M,
		\\
		\displaystyle
		1 \qquad \qquad \quad \qquad \quad O(t,\mathbf{x}) \leq O_{m}.
	\end{cases}	
	\label{phio_So}
\end{equation}
The term $N(O(t,\mathbf{x}))$ represents instead necrosis phenomena: viable cells indeed irreversibly acquire a necrotic fate when they experience a drop in the available oxygen concentration below to the basal level $O_{n}$. In this respect, $N$ in Eq.(\ref{eq:r}) reads as follows:
\begin{equation}\label{eq:necr_gg}
	N(O(t,\mathbf{x})) = \eta H(O_{n}-O(t,\mathbf{x})),
\end{equation}
where $H$ is again the Heaviside function and $\eta$ represents a transition rate. \\
\indent Finally, the term $T(u,O(t,\mathbf{x}))$ represents the radiobiological response of cells under the action of the treatment. To define it, we rely on the standard linear-quadratic (LQ) model, \cite{jones2001use}, which describes the surviving fraction $SF$ of cells in response to a single radiation dose. In our setting, cells mortality is defined following an innovative approach in the light of what we mentioned before i.e. in function of both (\textit{i}) oxygenation of the tissue and (\textit{ii}) intrinsic radiosensitivity of cell clones with respect to their epigenetic firm.  
Generally, the main parameters of LQ model $\alpha$ and $\beta$ are tissue specific coefficients, and we introduce 
variability in the action of radiotherapy  
according to the biological situations in which the therapy is applied, \cite{van2018alfa}. Specifically, we consider that the effectiveness of radiotherapy is related to hypoxia that affects therapeutic efficacy in a twofold way. A direct one by the fact that, as underlined before, oxygen is responsible for the enhancement of the detrimental effect of ionizing radiation which implies that radiotherapy is less efficient in the areas in which a lack of oxygen is observed. An inverse one by the fact that hypoxia selects for cells equipped by high resistance to hostile environments and low proliferative rates; this second characteristic makes cells, as already mentioned, intrinsically less exposed to radiotherapy action with respect the normal cells since the power of radiotherapy is to be able to damage the DNA consequently blocking the replication process.
In this light, we assume that the coefficients $\alpha$ and $\beta$ coefficients depend on introduce a variable $z$ which takes into account both the oxygen concentration (to simulate that the treatment is less effective in hypoxic areas) and the epigenetic characterization of cells (to
simulate that hypoxia-resistant cells are even less sensitive to radiotherapy). This concept is formalized describing $z$ as a product of two weights: (\textit{i}) the former, here named $w_u$, which depends on the epigenetic trait of the cell $u$ and (\textit{ii}) the second, here named $w_O$, from the oxygen concentration $O(t, s)$:
\begin{equation}
z(u,O(t,\mathbf{x}))=w_u(u) \cdot w_O(O(t,\mathbf{x})) \text{ where } w_u(u)=\frac{p_1(u)}{\gamma_{\textup{max}}}  \text{ and } w_O(O(t,\mathbf{x}))=p_2(O(t,\mathbf{x})),
 \label{eq:weights}
\end{equation}
to highlight the relation that exists between \textit{proliferation} and \textit{survival} (\textit{proliferation-survival trade-off}). In this light, inspired from the work presented in \cite{celora2021phenotypic}, we model the increasing dependence of $\alpha$ and $\beta$ parameters on the eco-evolutionary variable $z$ as:
\begin{equation}
 \alpha(z)=\alpha_{min}+(\alpha_{max}-\alpha_{min})z,
 \label{eq:alpha}
\end{equation}
and
\begin{equation}
 \beta(z)=\beta_{min}+(\beta_{max}-\beta_{min})z,
 \label{eq:beta}
\end{equation}
where $\alpha_{min},\alpha_{max},\beta_{min}\,\text{and}\,\beta_{max}$ are non-negative constants with $\alpha_{min}<\alpha_{max}$ and
$\beta_{min}<\beta_{max}$ which represent the maximum and minimum sensitivity to treatment (estimations of their values can be found in \cite{joiner}). Finally, we characterize cell mortality under the action of radiotherapy as follows:
\begin{equation}
	T(u,O(t,\mathbf{x}))= (\alpha(z)d+\beta(z)d^2)\cdot\delta_\text{times}(t)
	\label{eq:therapy}
\end{equation}
where: (\textit{i}) $\delta_{\text{times}}(t)=\sum_{T\in \text{times}}\delta_T(t)$, being $\delta$ the Dirac-delta, models the fact that the death factor due to therapy is only present during the administration time of the chosen protocol and (\textit{i}) $d$ is the administered dose.  
To complete the model, the viable cells turned into necrotic fate due to lack of oxygen, already described in Eq.\ref{eq:necr_gg}, are collected in the necrotic population, whose dynamic is described by:
\begin{equation}\label{eq:necrotic}
	\frac{\partial n}{\partial t}(t,\mathbf{x}) = N(O(t,\mathbf{x}))\rho(t,\mathbf{x}) = \eta H(O_{n}-O(t,\mathbf{x}))\rho(t,\mathbf{x}),
\end{equation}
\indent Switching on the molecular scale, in the same setting presented in \cite{chiarihypoxia}, the local concentration of oxygen is governed by a parabolic PDE where the spatially heterogeneous source term $V(\mathbf{x})$ captures the presence of intra-tumoural  {oxygen sources}. Moreover, oxygen diffuses within the tissue, naturally decays and it is consumed by viable cells. 
Its kinetics is described as follows:
\begin{equation}\label{eq:o}
	\frac{\partial O(t,\mathbf{x})}{\partial t} = \; \underbrace{\mu_{\textup{O}} \Delta_{\mathbf{x}} O(t,\mathbf{x})}_{\textup{diffusion}}\;-\;\underbrace{\lambda_{\textup{O}} O(t,\mathbf{x})}_{\textup{natural decay}} \;-\; \underbrace{\zeta_{\textup{O}} \rho(t,\mathbf{x}) O(t,\mathbf{x})}_{\substack{\textup{consumption by}\\ \textup{active tumour cells}}}+\underbrace{V(\mathbf{x})}_{\substack{\textup{inflow from}\\\textup{ {oxygen sources}}}},
\end{equation}
\noindent where $\lambda_{\textup{O}}$, $\mu_{\textup{O}}$ and $\zeta_{\textup{O}}$ are constant coefficients. We denote with ${\Upsilon}=\{(\textbf{v}_i,I_i)\in \Omega_s\times\mathbb{R}^{+}\}_{i=1}^{N_V}$ the set of  {oxygen sources} present in the tissue, where the $i$-th  {source} is defined by a couple in which the first element $\textbf{v}_i$ provides its the geometrical position and the second element $I_i$ is the rate of inflow of oxygen in the tissue \textit{via} it. Thus, the oxygen inflow can be described as a geometric source given by:
\begin{equation}
	V(\mathbf{x}) = \sum_{i=1}^{N_V} I_{i} \,e^{- \frac{(\mathbf{x}-{\mathbf v}_i)^2}{\sigma_{V}^2}},
	\label{eq:v}
\end{equation}
with $\sigma_{V}<<1$ to simulate a quasi point-wise source, coherently with the model presented in \cite{villa2021modeling}. In this respect, we specify that, in this modeling arrangement,  {oxygen source} characteristics,  {in time and geometry}, are time independent. 
 {We underline that the diffusion of oxygen within multicellular tumour spheroids typically occurs within a time frame of seconds. Consequently, employing a steady-state approximation is deemed acceptable, see for instance (\cite{grimes2014oxygen}). In our specific study, our objective was to investigate the success of radiotherapy, which, as previously mentioned, is strictly linked to the reoxygenation of the tumour microenvironment. Accordingly, we focused on the temporal changes in oxygen concentration within the same time window as cellular oxygen consumption.}

\subsection{Simulation details}
\label{subsec:Simulation details}
The spatial domain $\Omega_s$ represents a bi-dimensional section of a 4 cm-large tissue, i.e. $\Omega_s=[-2, 2]^2$ cm. The final observation time, is denoted by $t_{\textup{F}}$ and varies among simulations in correspondence of the relapse, identified as the time at which the total cell count reaches again the detection threshold  {($2.5\cdot 10^6$)} after the radiotherapy administration. 
\medskip\\
\noindent \textit{\underline{Initial and boundary conditions}}: Eq. \eqref{eq:a} that establishes cell dynamics, is equipped by the following initial condition:
\begin{linenomath}
	\begin{equation}
		\label{ica}
		a(0,\mathbf{x},u) = A \ \exp\bigg({-\frac{(\mathbf{x}-\mathbf{x}_C)^2}{2 \sigma_\mathbf{x}^2}-\frac{(u - u_C)^2}{2 \sigma_u^2}}\bigg), \,\, \quad \textup{for}\,\, \mathbf{x},u \, \in \, \Omega_s \times \Omega_p;
	\end{equation}
\end{linenomath}

\begin{linenomath}
	\begin{equation}
		\label{icn}
		n(0,\mathbf{x}) = 0, \quad \textup{for}\,\, \mathbf{x} \, \in \, \Omega_s,
	\end{equation}
\end{linenomath}
with $A>0 \quad \textup{s.t.}\,\, \rho(0,\mathbf{x})=\int_{\Omega_p} a(0,\mathbf{x},u) du <k$. The geometric point around which the cancer cell population is located at the initial time is denoted by $\mathbf{x}_C$; without loss of generality, we consider the case in which it is fixed at the center of the domain $\mathbf{x}_C = (0,0)$. Biologically, this setting reproduces, at the beginning of the numerical realizations, a node of malignant viable cells already present within the tissue, with the following characteristics: (i) each cell epigenetic state has a full Gaussian profile along the spatial dimension, centered at the starting point $\mathbf{x}_C$ and with a variance of $\sigma^2_\mathbf{x}= 0.008$ and (ii) the cell mass has a half-normal distribution in the trait space, with peak at  {$u_C=0$ (to capture the predominantly proliferative behavior that characterizes early-stage tumours) and with variance $\sigma^2_u= 0.08$. }
The initial cell configuration has a maximum value of $A=89.20$ cell/cm$^2$. In this respect, at $t=0$, the overall density $\rho$ of active individuals is symmetrically disposed w.r.t. $\mathbf{x}_C$ and, in normoxic condition, it is mainly composed of proliferative promoting cell variants with only a small fraction of survival promoting agents.\\
Eq. (\ref{eq:a}) has zero-flux conditions at the boundary of the epigenetic domain, i.e., $\partial_u a(\cdot, \cdot, 0)=\partial_u a(\cdot, \cdot, 1)=0$. This is consistent with the fact that malignant cells can not be characterized by a trait smaller than 0 or higher than 1. The same holds on the domain $\Omega_s$ under the assumption of considering the growth of the mass in a tissue slice where physical barriers as for instance bones, bounds of breast ducts or the lack of extra-cellular matrix, prevent the expansion of the mass out of them. 
\medskip\\
Turning on chemical kinetics, Eq. \eqref{eq:o} is completed with initial condition $O(0,\mathbf{x})$ that represents the steady-state of oxygen distribution in the tissue in absence of tumour cells with respect to different  {oxygen source} intensity whose value is specified case by case. We couple Eq. \eqref{eq:o} with zero-Dirichlet conditions at the boundary of the spatial domain $\Omega_s$ under the assumption of considering a sufficiently large tissue in which anoxic areas at the boundaries of the domain. 
In this respect, two geometrical layouts for  {oxygen sources} are adopted in this work: (i) one with a  {oxygen source} at the center of the domain, considering two inflow cases; we refer to them as $\Upsilon^{FV}=\{((0,0),\Tilde{I})\}$ ( {well oxygenated tissue, corresponding to a Full Vascularized layout - FV })  and $\Upsilon^{HV}=\{((0,0),0.5\Tilde{I})\}$ ( {poorly oxygenated tissue corresponding to a Half Vascularized layout - HV}) ; (ii) the other is a three ( {oxygen source})  layout, where all the sources are around the antibisector in the configuration  {$\Upsilon^{3V}=\{((-0.9,0.9),0.6\Tilde{I}),(1.0,-0.8),0.6\Tilde{I}),(0.8,-1.0),0.6\Tilde{I})\}$}. In the sequel we set the reference oxygen inflow $\Tilde{I}=1.57 \mu mol/ (cm^2 day)$.  {This value is estimated to ensure tissue oxygenation at physiological levels consistent with those observed in various tumour tissues (\cite{brown2004exploiting}). Specifically, it guarantees a distribution that ranges from sufficiently vascularized regions where oxygen levels is over the normoxic threshold (8 mmHg, as reported in \cite{brown2004exploiting}), to areas characterized by a significantly limited nutrient supply (below the maximum tolerated level of 1.5 mmHg, as described in \cite{brown2004exploiting}).}
\medskip\\
\noindent \textit{\underline{Parameters estimate}}:
The majority of model coefficients has a clear and direct biological meaning and therefore a proper estimate has been done taking advantage of the empirical literature. In this respect, we have referred to experimental works dealing with a wide spectrum of diseases since we here account for a generic tumour.  {From this perspective, the model configuration maintains a general nature. However, it is certainly feasible to parameterize the model for a specific tumour type if relevant data is available.}
The full parameters set up is listed in Table \ref{tab1}. \begin{table}[t!]
	\centering{
		\scriptsize
		\begin{tabular}{lllll}
			\multicolumn{1}{c}{\textbf{}} & \multicolumn{1}{c}{\textbf{Parameter}} & \multicolumn{1}{c}{\textbf{Description}} & \multicolumn{1}{c}{\textbf{Value [Units]}} & \multicolumn{1}{c}{\textbf{Reference(s)}}\\
			\hline
			\\
			\multirow{11}{*}{\rotatebox{90}{cell dynamics}} & $\mu_{p}$ & epigenetic variation rate & $8.64\cdot10^{-9}$ [day$^{-1}$] & \cite{doerfler2006dna} \\ 
			& $\mu_{s}$ &spatial diffusion rate & $ 3.11 \cdot 10^{-5}$ [cm$^2$/day] & \cite{martinez2012hypoxic} \\
			
			& $\gamma_{\textup{min}}$ &minimal cell duplication rate & $3.46\cdot 10^{-1}$ [day$^{-1}$] &\cite{martinez2012hypoxic} \\
			& $\gamma_{\textup{max}}$ &maximal cell duplication rate & $6.94\cdot 10^{-1}$ [day$^{-1}$] & \cite{martinez2012hypoxic} \\
			& $k$ &tissue carrying capacity & $10^6$ [cell/cm$^2$] & \cite{shashni2018size} \\
			& $\eta_O$ & oxygen selection gradient & 1 [day$^{-1}$] & model estimate \\
			& $\eta$ & rate of necrotic transition & 1 [day$^{-1}$] & model estimate \\
			&$\alpha_{min}$& Minimum $\alpha$ value for radiotherapy & 0.007 [Gy$^{-1}$]& \cite{joiner} \\
			&$\alpha_{max}$& Maximum $\alpha$ value for radiotherapy & 0.21 [Gy$^{-1}$]& \cite{joiner} \\
			&$\beta_{min}$& Minimum $\beta$ value for radiotherapy & 0.003 [Gy$^{-2}$]& \cite{joiner} \\
			&$\beta_{max}$& Maximum $\beta$ value for radiotherapy & 0.15 [Gy$^{-2}$]& \cite{joiner} \\
            \\\hline\\
			\multirow{8}{*}{\rotatebox{90}{oxygen kinetics}} & $\mu_{\textup{O}}$ & oxygen diffusion coefficient & $8.64\cdot 10^{-1}$ [cm$^2$/day] & \cite{martinez2012hypoxic} \\
			& $\lambda_{\textup{O}}$ & oxygen natural decay rate & $4.32 \cdot 10^{-1}$ [day$^{-1}$] & model estimate\\
			& $\alpha_{\textup{O}}$ &Michealis-Menten oxygen constant & $4.28 \cdot 10^{-9}$ [$\mu$mol/ cm$^2$] &\cite{dacsu2003theoretical} \\
			& $\zeta_{\textup{O}}$ &oxygen consumption rate & $2.25\cdot 10^{-6}$ [$\mu$mol/cell] & model estimate \\
			
			& $O_{n}$ & oxygen necrotic threshold & $0.7$ [mmHg] &\cite{brown2004exploiting} \\
			& $O_{m}$ & oxygen hypoxic threshold & $1.5$ [mmHg]  &\cite{brown2004exploiting} \\
			& $O_{M}$ & oxygen normoxic threshold & $8$ [mmHg] &\cite{brown2004exploiting} \\
			\\
			\hline
		\end{tabular}
		\caption{Reference parameters setting. 
 }\label{tab1}}
\end{table}
\medskip

\noindent \textit{\underline{Numerical method}}:
For the domain mesh and the implementation of the numerical resolution algorithm, a Python code is developed, using FEniCSx and Dolfinx packages, \cite{LangtangenLogg2017}. Specifically, we adopt an uniform discretization for the temporal and epigenetic domains and a triangular mesh with radial symmetry for the two-dimensional geometric domain. The system of partial differential equations is solved via a mixed solution scheme. We couple an explicit Euler method for the one-dimensional components of the domain (time and epigenetic trait) and a Galerkin finite element method for the dynamics on the geometric domain.  {In the latter case, we employ a weak formulation and consider an approximation with piecewise polynomials of first degree on the mesh elements and continuous over the whole domain. At each time step, we evaluate the reaction term $R$ in Eq. \ref{eq:a} and the consumption term $\zeta_O$ in Eq. \ref{eq:o} based on the previous time step. This linearizes the numerical problems and allows us to treat them symmetrically in the weak formulation. These constraints enable the code to simulate the problem with adequate accuracy, eliminating the need for extensive mesh refinements while maintaining reasonable execution times (in the order of minutes on a standard laptop).}

\subsection{Quantification of model results}
\label{subsec:Quantification of model results} 

Following our previous approach, \cite{chiarihypoxia}, to provide some qualitative indicators of tumour evolution and a more quantitative description of epigenetic trait distribution inside the mass, we divide the epigenetic domain $\Omega_p$ in three \textit{epigenetic bands}, denoted with L (low), M (medium) and H (high): $\Omega_p=\Omega_p^L\cup\Omega_p^M\cup\Omega_p^H$ with $\Omega_p^L=[0,0.3)$, $\Omega_p^M=[0.3,0.7]$, $\Omega_p^H=(0.7,1]$. 
We link the \textit{epigenetic bands} to the tissue regions in which they are the "optimal" ones in term of environmental selection \textit{via} the function $\varphi_o(O(t,\mathbf{x}))$, Eq. \eqref{phio_So}. We obtain a time-dependent partition of the spatial domain 
$\Omega_{s}(t)=\Omega_{s}^L(t)\cup\Omega_{s}^M(t)\cup\Omega_{s}^H(t)\cup\Omega_{s}^N(t), \,\, \forall t \in T $ where $L$, $M$, $H$ superscripts correspond to the areas of the domain where the fittest epigenetic trait belongs to the correspondent epigenetic band and $\Omega_{s}^N(t)$ is the necrotic area (oxigen below $O_n$).\\
To analyze the results of our simulations, we use local and global indicators.\\
In order to spatially characterize the tumour mass, we take into account the local number density $\rho(t,x)$ already presented in Section \ref{sec:Materials and methods} and we introduce the \textit{band-specific local number densities}:
\begin{equation}
	\label{eq:rho_I} 
	\rho_j(t,\mathbf{x})=\int_{\Omega^I_p}a(t,\mathbf{x},u)\, du \qquad \text{for } j\in\{L,M,H\}
\end{equation}
To give some global indicators and analyze their evolution in time we introduce:
\begin{itemize}
 \item the \textit{total cell count}, providing the size of the tumour population:
 \begin{equation}
	\label{eq:cell_count} 
	\Gamma(t)=\int_{\Omega_s}\rho(t,\mathbf{x})\, d\mathbf{x}
 \end{equation}
 \item the \textit{average epigenetic trait}, providing a representation of the epigenetic spectrum present in the mass: 
 	\begin{equation}\label{eq:aveptr}
 	 \bar{g}(t) = \frac{1}{\Gamma(t)} \int_{\Omega_p} u\,\bigg( \int_{\Omega_s} a(t,x,u)\, d\mathbf{x}\bigg)\, du,
 	\end{equation} 
 \item the \textit{average radio-sensitivity index}:
	\begin{equation}\label{eq:rsindex}
		\bar{\alpha}(t) = \frac{1}{\Gamma(t)} \int_{\Omega_p}\int_{\Omega_s} \alpha(z(u,O(t,\mathbf{x})))\, a(t,x,u)\, d\mathbf{x}\, du,
	\end{equation} 
\end{itemize}
With respect to the radiotherapy parameters, we here use $\alpha$ as a qualitative indicator since we are interested in the relative variation with respect to minimum and maximum values and $\beta$ has the same dependence on the $z$ function, just rescaled in its range, Eqs. \eqref{eq:alpha}, \eqref{eq:beta}.\\ 
Finally, we introduce a global index for the environment description, which is the \textit{oxygen total amount}
\begin{equation}
 O(t) = \int_{\Omega_s}O(t,\mathbf{x})\, d\mathbf{x} \, . 
\end{equation}

\section{Results}
\label{sec:Results}
The rationale of this work is to investigate how differences in the tumour radiotherapy response could be related to the spatially heterogeneous distribution of intratumoural blood vessels in tumour tissues. One of the factor that may lead to therapeutic failures is the development of intra- and inter-patients resistance. Different ecological niches, in terms of vessels potency and relative dispersal, lead to te selection of cells with different characteristics that may pave the way to the emergence of therapeutic resistance. In these veins, the resulting heterogeneous tumour microenvironment is of clinical interest to find the optimum \textit{patient-specific} protocol.\\
To investigate this phenomenon, according with the approach settled in our previous work \cite{chiarihypoxia}, we firstly focus on studying these evolutionary dynamics in a relatively simple setting choosing  {a mass formed by tumour cells that grows close to an oxygen source}.  Specifically, we are interested in two different environmental conditions, designed to represent an high and low efficient vascular network, respectively, finally comparing them in terms of therapy efficacy.\\
In both cases, we assume that the treatment is applied only in presence of a sufficient highly concentrated tumour mass, to reflect that, to be treated, masses have to be visible via diagnostic imaging. Specifically, we set a detecting threshold in correspondence of masses constituted by $4\cdot 10^6$ cells. With respect to the applied protocol, we focus on one of the most common in conventional clinical practice, according to which, patients receive the same radiation dose in all the different subregions of the tumour volume, following a fractionation schedule that provides, for a dose of approximately 2 Gy (Gray) delivered once a day, Monday to Friday, up to a total of 50-70 Gy. Specifically, in our case, the therapy ends up in 6 weeks for a total dose amount of 60 Gy. The mass evolution is described in terms of numerosity, morphology and epigenetic characterization, investigating the relation between these different aspects in \textit{repopulation} phenomena. Moreover, special emphasis is devoted to tumour-host interaction, looking to oxygen dynamics in relation to the \textit{radiosensitivity} of the mass considered.

\subsection{Case 1 -  {Well oxygenated tissue}}
\label{subsec:Case 1 - highly efficient single vessel}
\begin{figure}
\includegraphics[width=1\linewidth]{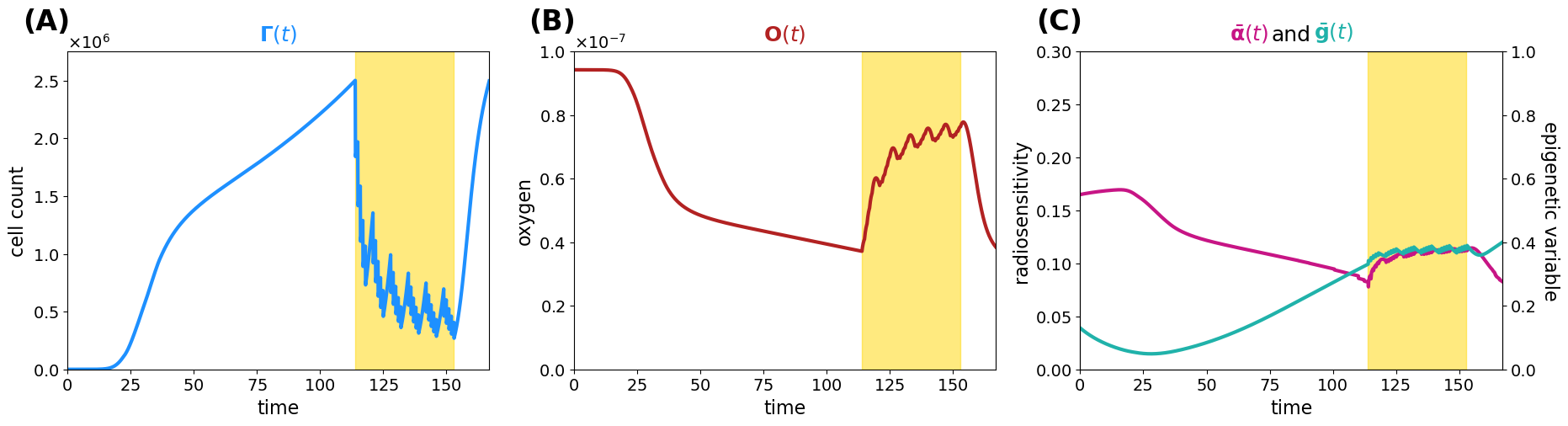}
\includegraphics[width=1\linewidth]{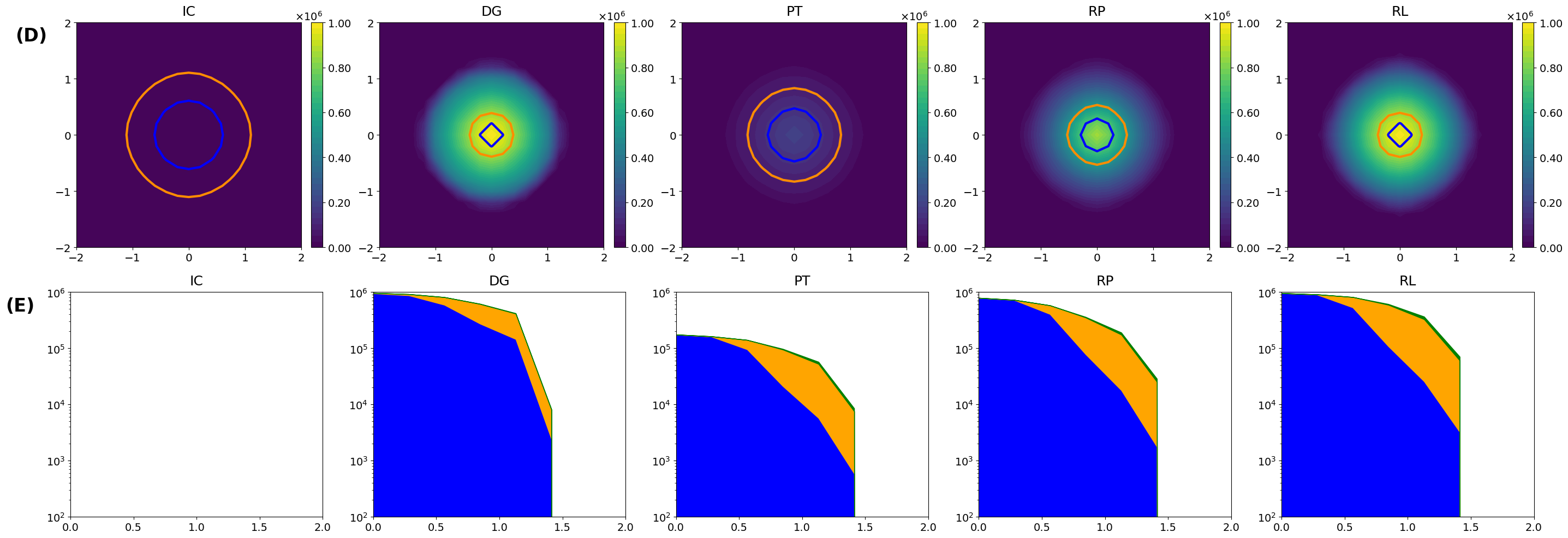}
\caption{Simulation of tumour mass, growing from the center of the domain,  {in the case of a well oxygenated tissue}, $\Upsilon^{FV}$.layout (A) Global cell count evolution. (B) Total oxygen amount evolution. (C) Average radiosensitivity index and average epigenetic trait evolution. (D) $\rho(t,\mathbf{x})$ for $t$ correspondent to  {tumour onset (IC),} diagnosis (DG), post-treatment (PT), repopulation (RP) and relapse (RL) times. Contour lines detect the optimal areas for high (green), medium (orange) and low (blue) epigenetic bands  {(note that here green line is not present as it would divide the optimal area for high epigenetic band from the necrotic area, which is not present with this oxygen inflow).} (E) Slice of $\rho(t,\mathbf{x})$ taken on the positive bisector in logarithmic scale. The edge represents the total amount $\rho(t,\mathbf{x})$ while  {the edges of colored areas represent the incremental densities, providing a visual indicator for the geometrical characterization of the epigenetic composition of the mass (blue for low, orange for medium and green for high epigenetic band).}}

\label{fig:simu1}
\end{figure}

In the first simulation setting, Figure \ref{fig:simu1}, we observe the growth of a tumour mass in a sufficiently oxygenated environment provided by $\Upsilon^{FV}$ layout. Looking at the evolution of its total cells count $\Gamma(t)$, panel (A) of Fig. \ref{fig:simu1}, the  {efficient vasculature of the tissue provides a sufficient nutrient supply}, leading to a rapid evolution of the mass. The malignant cells number shows a Gompertz-like profile, coherently with the biological evidences, see for instance \cite{folkman1973self,oraiopoulou2017vitro}. 
Assuming to be able to diagnose the tumour burden, as soon as, the mass approaches the detection threshold, radiotherapy is applied, according to the above mentioned protocol, leading to  {a tumour reduction of nearly the 90$\%$ of its volume at the diagnosis}. The shape of the curve, week by week, fits the decreasing strength of the treatment in time, showing a strong efficacy in the first phases that gradually reduces in correspondence of a smaller portion of cells that could be hit. As it can be observed, once that the treatment protocol is completed, the failed eradication allows to a quick relapse of the mass that restarts to grow even faster than the settlement phase; a restored bulk of the same dimension of the one before the clinical intervention could be observed at the end of the simulation.\\
The three-phase \textit{expansion-contraction-expansion} dynamics is even more highlighted looking at the evolution of the local density $\rho(t,x)$, represented in panel  {(D)} of Fig. \ref{fig:simu1}. Specifically, we focus on observing its profile at  {five} representative time points:  {\textit{i)} the mass onset (IC), \textit{ii)} the tumour detection (DG - diagnosis), \textit{iii)} the end of the treatment (PT - post treatment), \textit{iv)} during the tumour regrowth (RP - repopulation) and \textit{v)} the final configuration (RL - relapse)}. As it can be observed, the mass develops almost-radially around  {the more oxygenated area and no differences in terms of shape could be highlighted comparing the second and the last panel}. This is coherent with the fact that the tumour invasion is the result of the proliferative and the diffusive potential, with respect the environmental conditions faced. Interestingly, at the end of radiotherapy administration, the cell density of the mass is underneath the detection threshold, see panel (D)-PT of Fig. \ref{fig:simu1}. 
However, looking to the corresponding DG and PT panels in panel ( {E}) of Fig. \ref{fig:simu1}, where the tumour densities of the three different \textit{epigenetic bands} $\rho_j\,\,\text{for}\,\,j=L,M,H$, are locally summed in space, the radiotherapy reduction corresponds to approximately the  {$80\%$ of the tumour bulk, switching in number from $10^6$ to $2 \cdot 10^5$ while the edges remain almost constant in number}. The remaining small node of cells, constituted by the  {$10\%$} of the total original mass, is the responsible of tumour relapse, in accordance with the \textit{repopulation} phenomena. The rapid growth of the mass observed could be reasonably linked to tumour-host interaction and in particular to the \textit{reoxygenation} after the radiotherapy. From a local point of view, this can be seen analyzing the contour lines at different time steps in panel (D) of Fig. \ref{fig:simu1}, that detect the  {extension of the} optimal areas for the three epigenetic bands.  {Blue circle indices the area in which more proliferative epigenetic traits are favourite to colonize the tissue, the orange the one for the medium-resistant cells and the green (here not represented) for the high-resistant cells ones. Notice that the absence of a boundary of the optimal area for more hypoxic cells denotes the fact that in the entire tumour tissue the amount of oxygen is sufficiently high to avoid necrosis so high hypoxic-resistant cells could potentially fully colonize the tissue slice.} The areas enclosed by the contour are wider after therapy (PT plot), due to the lowering of oxygen consumption as a consequence of the elimination of the tumour mass by the radiation administration.\\
From a global point of view, looking indeed to the evolution in time of total oxygen amount available in the tissue slice, panel (B) of Fig. \ref{fig:simu1}, an increase in terms of its concentration occurs during the treatment administration. The rationale, as already anticipated in the Introduction, is that the reduction in number of tumour cells during the treatment is reflected in a minor consumption of the available nutrients, leading to restored oxygen delivery in the tissue that was, in the early development phases, compromised due to the high metabolic requirement. As we can see, at the end of the radiotherapy protocol, the survived tumour cells could completely exploit the available oxygen, leveraging all their duplication potential and leading to a quick mass regrowth.\\
In eco-evolutionary terms, it is interesting to investigate how the \textit{cooperation} and/or \textit{out-competition} phenomena could potentially influence this dynamics. In particular, as shown in our previous work \cite{chiarihypoxia}, the selection dynamics occurring in the tumour-host interaction could carry out a crucial role in terms of tumour aggressiveness and treatments could act, in terms of environmental stressors, as bottlenecks that fuel this dynamics. Three aspects are of particular interest from our point of view: \textit{i}) how, the pre-treatment history of a tumour could impact on the radiosensitivity of the mass at the beginning of treatment delivery; \textit{ii}) the effect of radiotherapy as a bottleneck selecting for resistant epigenetic traits and finally \textit{iii}) how the consequences of the treatment action on tumour microenvironment could impact in the future of tumour mass.\\
In this direction, we focus on investigating the evolution in time of the global average epigenetic expression of the mass $\bar{g}(t)$, Eq. \eqref{eq:rsindex} and the corresponding one of the radiosensitivity index $\bar{\alpha}(t)$, Eq. \eqref{eq:aveptr}. Analyzing the $\bar{g}$ profile in time, we can observe that the high oxygenated environment that characterizes the early phases of tumour growth leads to an initial deflection of the average epigenetic expression. Indeed in proximity of  {the more oxygenated area}, proliferating cells have a strong evolutionary advantage with respect to all the other epigenetic traits present in the mass, out-competing them and becoming the predominant ones in the mass. As the mass expands, conquering less oxygenated regions, the action of natural selection slowly leads to the emergence of more resistant epigenetic traits, mildly shifting the tumour towards resistance to hypoxia development. However, a strong epigenetic switch in terms of average composition of the mass, during and after the treatment, is not remarkably highlighted. Indeed, under these specific environmental conditions, the growth of the mass is only slightly affected by natural selection in the short time interval before detection, and thus it is mainly composed by proliferating cells  {and a minor part of medium resistant cells}. \textit{Reoxigenation} after the treatment gives a further advantage to the proliferating cells. 
This is confirmed looking at the epigenetic composition in terms of specific epigenetic bands densities represented in panel (E) of Fig. \ref{fig:simu1}. In this light, $\bar{g}(t)$, coherently, remains pretty constant with only small fluctuations,  {during the treatment}. The jagged $\bar{g}(t)$ profile is due to the fact that: on one hand, during the resting days of the treatment, in accordance with resources availability, proliferating cells quickly \textit{repopulate} the tumour mass; on the other hand, during the effective treatment days, the higher effect of radiotherapy on proliferating cells leads to their decrease. \\
A more dynamic profile characterizes the evolution in time of the \textit{radiosensitivity} of our mass. As we can observe, in the early expansion phases, despite the promising tumour composition in terms of radiotherapy efficacy, the radiotherapy index rapidly decreases. This is due to the fact that the growing mass approaches less oxygenated regions due to both an increased oxygen consumption by the mass and the further distance from the  {high oxygenated area} and thus the tumour results to be less sensible to radiotherapy administration, independently by its epigenetic characterization. In contrast, the \textit{reoxygenation} phenomenon observed during the radiotherapy leads instead to an increasing responsiveness of mass to the treatment. The jagged profile of $\bar{\alpha}(t)$ is less regular with respect the one observed for the average epigenetic index $\bar{g}(t)$, and, in this respect, our results suggest that the heterogeneity in terms of radiotherapy efficacy are mostly oxygen-driven with a minor role of epigenetic mass composition. 
Unfortunately, coherently with the experimental evidences, \cite{hong2016tumor}, the benefits of radiotherapy in terms of \textit{reoxigenation} and thus, the increase of \textit{radiosensitivity} of the mass has only a temporary effect: in a quick time window, the tumour mass re-acquires its scarce sensibility, being able to conquer and to survive in hypoxic regions where oxygen concentration is high enough to allow for clonogenic survival, but it is low enough to protect tumour cells from the effects of ionizing radiation.

\subsection{Case 2 -  {Poorly oxygenated tissue}}
\label{subsec:Case 2 - inefficient single vessel}
\begin{figure}
\includegraphics[width=1\linewidth]{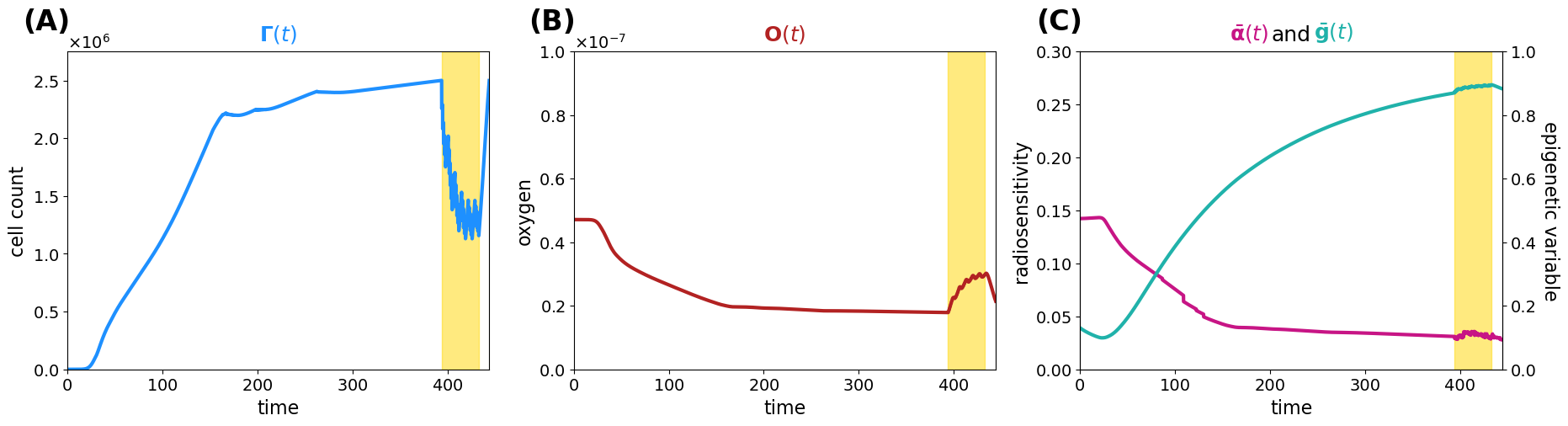}
\includegraphics[width=1\linewidth]{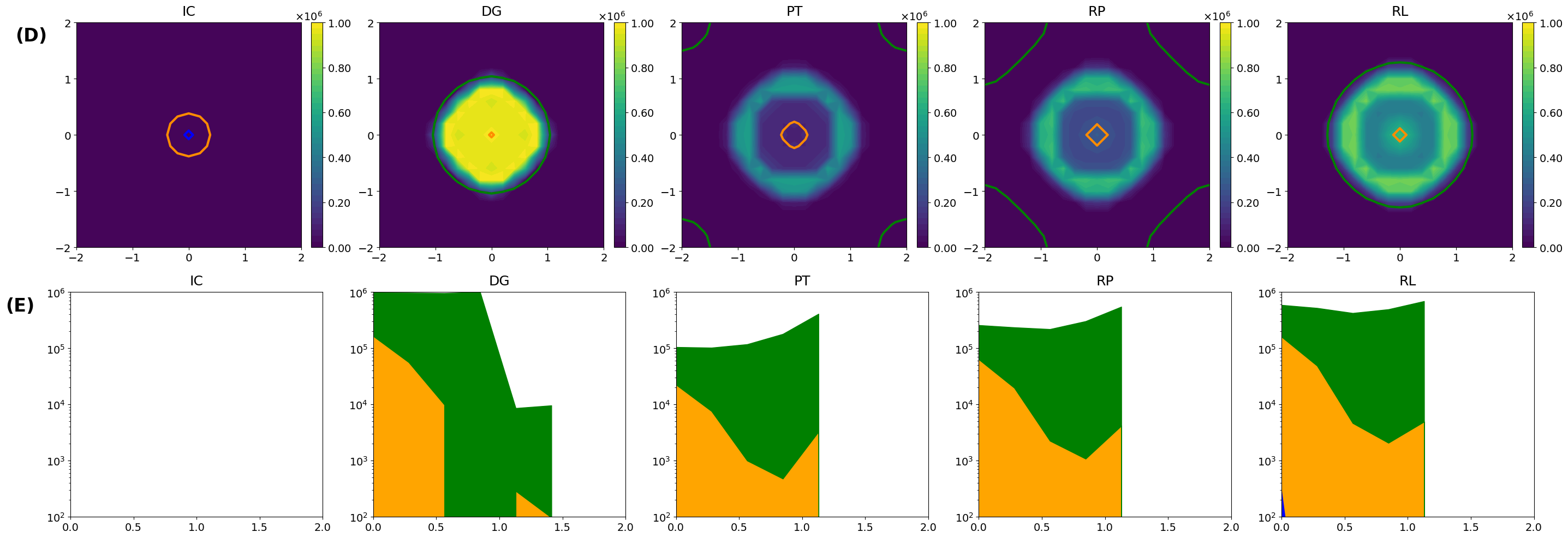}
\caption{Simulation of tumour mass, growing from the center of the domain, with oxygen  {levels in accordance with the $\Upsilon^{HV}$ layout}. 
Plots description as in caption of Fig. \ref{fig:simu1}.
}
\label{fig:simu2}
\end{figure}

\begin{figure}
\includegraphics[width=1\linewidth]{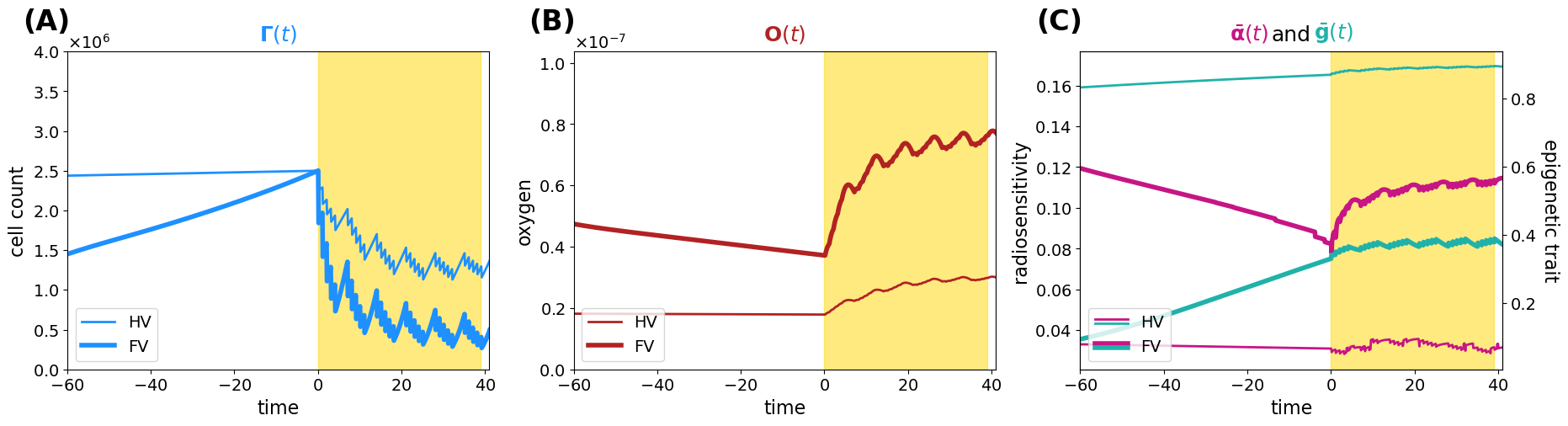}
\caption{Comparison between case 1 and case 2. The figure shows an overlap of (A), (B), (C) plots presented in Fig. \ref{fig:simu1} and Fig. \ref{fig:simu2}). The time axis has been rescaled so that time zero coincides with the begin of the therapy for both the experiments and plots are zoomed around it.}

\label{fig:comp}
\end{figure}

In the second simulation setting, we observe the growth of a tumour mass in a harsher environment as consequence of an inefficient  {nutrients supply}. In this perspective, we consider  {a tissue slice in which the oxygen inflow by the micro-vasculature is halved with respect to the previous case and we analyse the same eco-evolutionary features.}
Looking at the results shown in Figure \ref{fig:simu2}, marked differences can be observed on tumour evolution with respect to the previous case.\\ 
First of all, the evolution of tumour cells total count $\Gamma(t)$ is significantly different (panel (A) of Fig. \ref{fig:simu2}); the smaller amount of available resources slows down the tumour development, leading to a 
mass that needs approximately  {more than three} times more to reach the detecting threshold. Its profile moreover reveals that the radiotherapy protocol is less effective with respect to the previous case; the mass reduces again but a more dense nidus of resistant cells, that amounts to over the  {40}$\%$ of the total volume, is not eradicated. Additionally, 
once that the treatment protocol is completed, the failed eradication allows to a quick relapse of the mass; the restored bulk at the end of the simulation is indeed morphologically identical to the one before the clinical intervention but it reaches the same volume, in a shorter time-window with respect the early stages of the mass growth. The characteristic three-phase \textit{expansion-contraction-expansion} dynamics is again observed, however it is affected by completely different mechanisms with respect the previous case as it is shown by the rest of the eco-evolutionary dynamics.\\ 
The evolution in time of the tumour density $\rho$ reveals, in contrast with the previous scenario, a moderate dense ring of resistant cells still present at the end of treatment protocol. Its presence naturally affects the re-growth of the mass; and, as it could be observed in panel ( {D})-RP of Fig. \ref{fig:simu2}, two simultaneous dynamics could be detected: \textit{i}) a \textit{repopulation} of the tumour region that starts from the center of the mass in proximity of the  {more oxygenated tissue area}, suggesting that, also in this case, a silent mass is not eradicated coupled with \textit{ii}) an uninterrupted expansion of the rim of resistant cells.\\
This macroscopic difference between the two cases strongly depends on the different  
interactions between the tumour and the microenvironment that lead to completely different \textit{radiosensitivity} and, consequently, different \textit{repopulation} and \textit{reoxygenation} dynamics in the mass. 
In particular, two are the aspects that have to be considered to identify the underpinning dynamics: \textit{i}) a less oxygenated environment 
could lead to the emergence of intrinsically more resistant epigenetic traits; \textit{ii}) an inefficient  {tissue oxygenation} naturally suppresses the radiotherapy efficacy.\\
Focusing on the first aspect, coherently with what we showed in our previous work \cite{chiarihypoxia}, in the early stages of mass development, the tissue colonization results from the cooperative relations between different specialized cell variants, enhancing the importance of epigenetic composition on tumour development. Analysing the local number densities of the different sub-groups $\rho_j$ for $j=L,M,H$, it can be noticed that the tumour composition is affected by the spatial variability of oxygen concentration and environmental gradients lead to the selection for cells with epigenetic characteristics that vary  {in accordance with the oxygen gradient}. 
In particular, cells characterized by medium and high resistance colonize the mass in percentages that increase moving far away from the nutrient source. This emerging dynamics suggests the development of the classical ring structure that characterizes spheroids; in this specific case, the mass is evolving  {developing an inner core of medium resistant epigenetic traits surrounded by a rim of high resistant cells}. A mixed composition of this type shapes the radiosensitivity of the mass and results in an heterogeneous response with respect the cellular subtypes involved. The reduction in tumour burden provided by the action of the therapy is  {comparable with respect the previous case. Differently, the tumour edges remain almost insensible to the treatment, increasing in number.} These results highlight how in tumour-host interaction, the selection dynamics could carry out a crucial role in terms of tumour aggressiveness and that treatments could act, as environmental stressors and bottlenecks that fuel this dynamics. In particular, looking at panels DG and PT of (D) plots of Fig. \ref{fig:simu2}, two aspects of particular interest naturally emerge: \textit{i}) the impact of pre-treatment history of a tumour could impact on the radiosensitivity of the mass; \textit{ii}) the role of radiotherapy as a bottleneck selecting for resistant epigenetic traits.\\
In this perspective, the residual ring that emerges at the end of the treatment is the result of the presence, in the region of interest, of pre-existing more resistant epigenetic traits. This is perfectly in line with the biological hypothesis that the action of radiotherapy is affected by the tissue oxygenation level in a two-fold way: \textit{i}) low oxygenation levels promote the emergence of radio-resistant cells and we will refer to it as \textit{epigenetic-driven resistance}; \textit{ii}) oxygen is fundamental to fix on DNA of the cells the damage induced by radiation and we will refer to it as \textit{purely oxygen-driven resistance}.\\
Notably, also in this second case, \textit{reoxygenation} could be observed as a consequence of tumour cells killing by radiotherapy, but its restoring dynamics is characterized by a slower slope with respect to the previous case being. The underpinning reason relays in tumour mass composition: the presence of an heterogeneous mass with medium and high resistant cells implies that not all the viable cells are killed by the treatment and, thus that the oxygen consumption does not stop during the treatment. Despite the slower dynamics, \textit{reoxygentation} fuels more remarkably the tumour \textit{repopulation}. The larger nidus of survived cells are indeed strongly advantaged by the presence of new available resources, favoring their proliferation. It is interesting to notice how the \textit{repopulation} of the mass is the result of \textit{cooperation} phenomena in which  {medium proliferating cells  colonizing the inner region of the tumour tissue; as well as harsher regions are instead repopulated by increasing resistant epigenetic traits}. This phenomenon is highlighted in panel (D-RL) of Fig. \ref{fig:simu2} in which the ring structure, previously sketched, definitively emerges; an entire rim  {predominantly} constituted by high resistant epigenetic traits bounds indeed the tumour mass.\\
The eco-evolutionary indexes that we are considering, the average epigenetic trait, Eq. \eqref{eq:aveptr}, and the radio-sensitivity index, Eq. \eqref{eq:alpha}, are able to reveal additional interesting information. Looking at the average epigenetic trait evolution, the interaction with an harsh environment strongly forces the hypoxia-resistance development trend. The epigenetic shift that is observed rapidly converges towards resistant epigenetic traits in the class of medium and high resistant cells and thus the mass is constituted in higher percentage by these cellular subtypes. Its profile moreover highlights: \textit{i}) the strength of the selective bottleneck induced by the action of the treatment that provides a further shift towards an increasing radio-resistance and \textit{ii}) the effect of \textit{reoxygenation} on \textit{radiosensivity} of the mass as confirmed by the deflection in the immediate time-window after the treatment. The last information could be exploited from the therapeutic point of view being a mass sensible to proliferation targeting approaches. In the same veins the evolution of the radio-sensitivity index shows a decreasing profile during all the time-window of observation. This sharp trend is due to the harsh environment in terms of oxygenation:
hypoxia, already by itself, constitutes a valid element to decrease radio-sensitivity of the mass which is affected by oxygen deprivation also in an indirect way via the selection of resistant epigenetic traits. 
Moreover, the radio-sensitivity index shows a more irregular dynamics with respect the previous case, coherently with the fact that in this case the radiotherapy response is guided by both the oxygenation levels and its epigenetic characterization.
\\
To summarize, a comparison of the dynamics is plotted in Figure \ref{fig:comp}; bold and thin lines refer respectively to low and high oxygenated microenvironment. Interestingly, our results suggest that nominal tumour size alone is insufficient to predict growth dynamics and that the definition of personalized indexes is needed to define an efficient therapeutic plan. In fact, the time evolution of the total cell counts, panel (A) of Figure \ref{fig:comp}, reveals how two patients with a similar tumour volume could have a distinct tumour–host co-evolution history, which results in different responses to the same radiotherapy protocol, coherently with the clinically observed inter-patient variability in terms of therapy response. Furthermore, a particular behaviour is highlighted: tumour volumes close to their carrying capacity, in terms of the maximum tumour extension that can be reached with respect to the available resources, result to be less sensitive to radiation-induced damage (low oxygen case); on the other hand, tumours far from their carrying capacity are instead more sensitive to the radiation (high oxygen case). In this light, our results are interestingly in line with the ones presented in \cite{poleszczuk2018predicting}, in which a patient-specific index, named the Proliferation Saturation Index (PSI), estimated on patient data, is introduced as predictive tool of the radiotherapy response. In their work, the authors hypothesize that tumour characterized by an high PSI, i.e. close to their carrying capacity, are composed by only a small proportion of proliferating cells highly sensitive to radiation-induced damage and thus, a less effective therapeutic impact could be expected. Our results confirm this hypothesis, further revealing the dynamics that determine the differences in tumour composition as result of the heterogeneity from patient to patient in tumour microenvironment. In this respect, see panel (B) of Fig. \ref{fig:comp}, our findings moreover suggest that, even if the restoring oxygenation is, in percentage terms, the same in the two cases (close to the  {50}$\%$), a completely different dynamics is observed. Indeed, in the first case (high oxygen level), looking at the in-time evolution of the radiosensitivity $\bar{\alpha}$ in panel (C) of Fig. \ref{fig:comp}, the radio-induced reoxygenation is sufficient to prompt the mass towards a higher sensibility to the treatment; at the contrary, in the case of low oxygenation, a degenerating dynamics towards a more resistant mass is observed, even in presence of higher quantity of available oxygen. In this view, our approach enriches the number of information that can be harvested, providing a platform that target both the microenvironment and the epigenetic characteristics of the mass. Therefore, it can potentially used to make predictions based on the state of the mass once discovered and to guide the protocol choice. 
 {Indeed introducing a mathematical framework which describes the cell behavior through phenotypic characterization gives the chance to use theoretical information about non-observable phenomena to explain the emergence of dynamics observed at the macroscopic level in terms of tumour expansion or reduction in the presence of treatment. Quantitative comparisons could be performed using data such as the ones reported in \cite{zakelj2019electrochemotherapy}, where the number of cells in each epigenetic band is computed and the tumour mass composition is identified in terms of percentage. Nevertheless, although there is increasing evidence of the existence of a continuum spectrum of phenotypes (\cite{alfonso2019modeling}), it is more challenging to obtain data in this regard. }
\subsection{ {Tailored radiotherapy protocols}}
\label{subsec:Dose painting}
\begin{figure}
\includegraphics[width=1\linewidth]{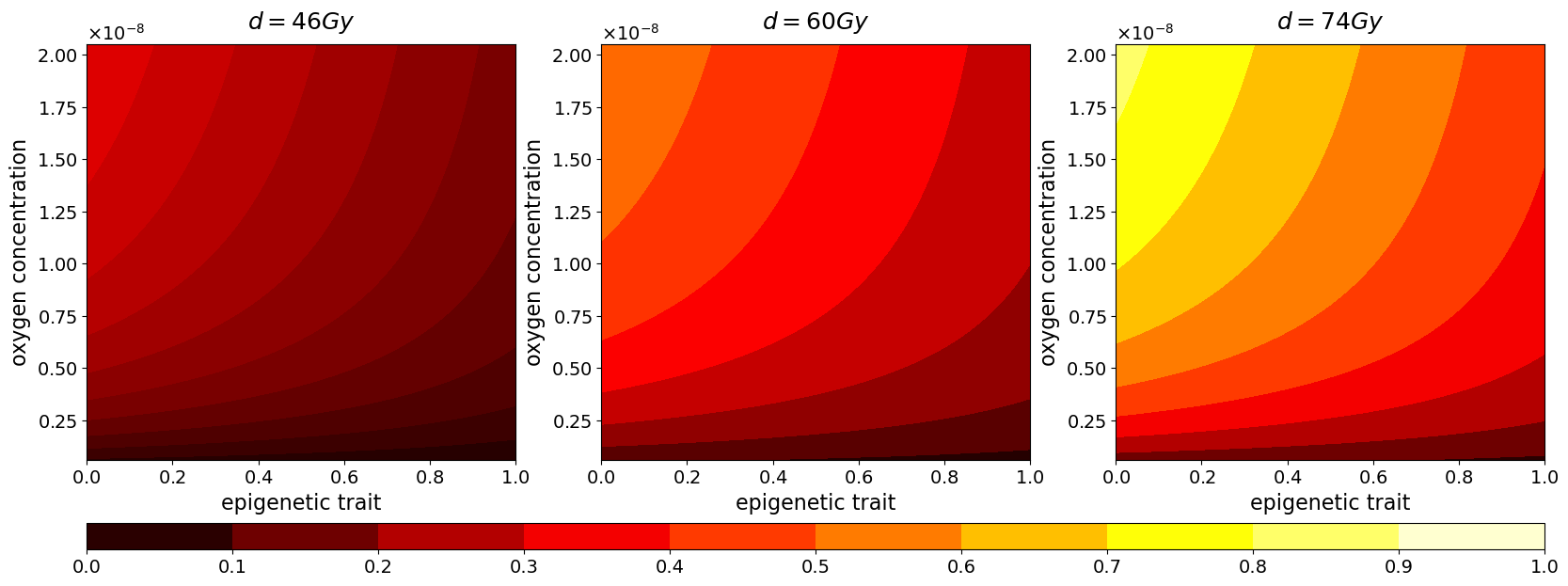}
\caption{Normalized radiotherapy effectiveness depending on epigenetic trait (x-axis) and oxygen density (y-axis) in the cases of a total dose of $46$, $60$, $74$ Gy.}

\label{fig:radio_efficacy}
\end{figure}

\begin{figure}
\includegraphics[width=1\linewidth]{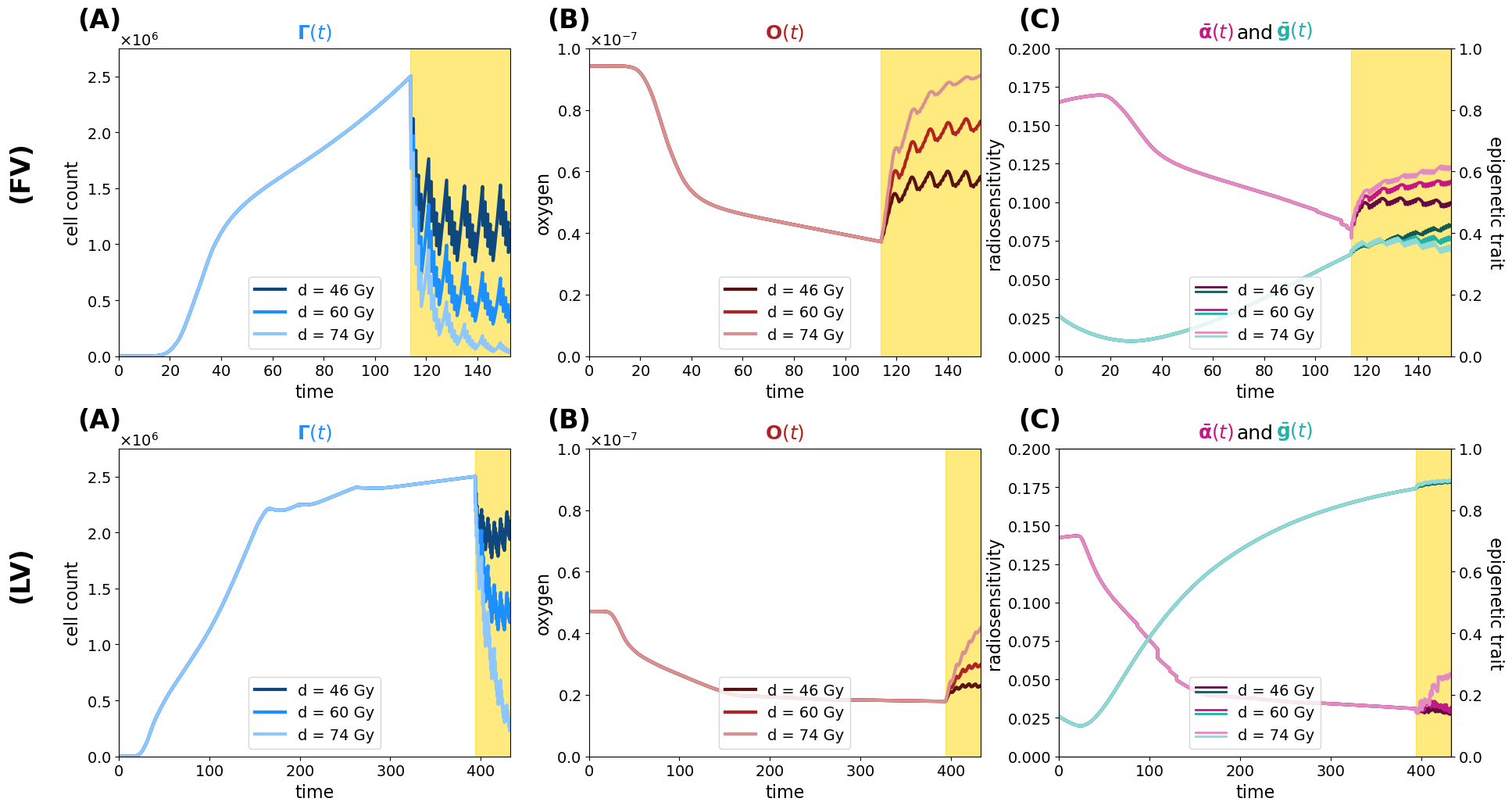}
\caption{Simulation of tumour mass, growing starting from the center of the domain,  {in the two oxygenation conditions, the well and the poorly oxygenated tissue slices which correspond to the} $\Upsilon^{FV}$ layout (first row) and $\Upsilon^{HV}$ layout $\Upsilon^{HV}$ (second row). In each row, plot descriptions follow Fig.\ref{fig:simu1} caption. Different colors depict experiments in which different total amounts of radiation are used ($46$, $60$ and $74$ Gy as indicated in legends).}
\label{fig:diff_dosage}
\end{figure}

In conventional clinical practice, most patients treated with radiotherapy receive a similar dose and fractionation scheme. In particular, at present, the same radiation dose is delivered to all subregions of the tumour volume, regardless of their individual biology and radiosensitivity. As shown, oxygen concentration can greatly modify the patient response; in this light, new treatment modalities such as Intensity-Modulated RadioTherapy (IMRT) have emerged, aiming to modulate the delivered dose over small volumes that are distinguished with respect the oxygenation level, see e.g.\cite{ling2000towards,baumann2016radiation}. This customization of radiotherapy, based on spatial information drawn from hypoxia imaging, is generally known as \textit{dose painting} and, in principle, it consists in delivery selective boosting dose to radio-resistant regions, \cite{grimes2017hypoxia}. However, to fully exploit the strength of these new techniques, hypoxia levels information needs to be coupled with the knowledge of tumour composition in terms of therapy resistance that, as mentioned, strongly impacts treatment efficacy. Thus, detailed information about the internal structure of the tumour in terms of epigenetic traits and phenotypes is required to define the best radiation dosimetry plan.\\
In these veins, we handle our modeling approach to investigate the dose-efficacy relationship with dose escalation, to suggest the optimal total dosage while reducing treatment-induced toxicity; specifically, our aim is to explore how a different prescription of radiation in regions characterized by higher hypoxia may or may not affect the success of treatment. To do this, we compare our previous results with two additional radiotherapy protocols that differ from the previous one in terms of the total radiation dose administered; in particular, we study the effect of a lower and a higher dosage compared to the previous case ($46$ total Gy and $74$ total Gy vs. $60$ Gy, respectively) while maintaining once-daily administration from Monday to Friday for $6$ weeks. Figure \ref{fig:radio_efficacy} shows a representative indication of the differences in efficacy for the three different scenarios, in relation to tumour microenvironment oxygenation and epigenetic composition of the mass. The results of the experiment in terms of \textit{repopulation}, \textit{reoxygenation}, and \textit{radiosensitivity} are shown in Figure \ref{fig:diff_dosage}: the first row (FV) refers to the case of high oxygenation, while the second row (HV) refers to the case of low oxygenation. The color code indicates that lighter colors correspond to higher total doses.\\
As we can see in both cases, the effectiveness of the treatment is strictly dependent on the dose administered. In fact, higher doses correspond to smaller portions of the tumour that can survive, although, as previously mentioned, it is observed that, at the same dosage, radiation therapy is less efficient in the case of low oxygenation, compare panels (A-FV) and (A-HV) of Fig. \ref{fig:diff_dosage}.\\
Regarding \textit{reoxygenation} phenomenon, both panels (B-FV) and (B-HV) of Fig. of Fig. \ref{fig:diff_dosage} reveal that, in both low and high hypoxia case, the higher is the dosage, the more effective is the tissue \textit{reoxygenation}. In the case of high oxygenated tissues, the epigenetic composition that characterizes a mass under this condition, shown in panel (D) of Fig. \ref{fig:simu1}, 
is mainly composed by proliferating cells at the time of treatment administration; therefore, as expected, a lower dosage of radiotherapy implies a lower percentage of destruction of highly sensitive cells. Comparing the two cases with different oxygenation, (FV) and (HV), the different speed at which the tissue reoxygenates in all three tested dosages is consistent with the selection phenomenon mentioned above. We indeed expect that, as indicated by the radiosensitivity index and by the average epigenetic composition in the case of low oxygenation (shown in plots (C-FV) and (C-HV) of Fig. \ref{fig:diff_dosage}),  
there will be a smaller portion of radiosensitive cells, therefore a smaller portion of cells killed and consequently a slower reoxygenation, due to higher oxygen consumption.\\
The dynamics revealed by the model in terms of \textit{radiosensitivity} and epigenetic composition of the mass, as a function of the dose amount, are interesting from a therapeutic point of view. Indeed, looking at the results shown in panel (C) of Fig. \ref{fig:diff_dosage},  
in presence of high oxygenation,  {it is evident that at higher doses, the likelihood of killing resistant cells becomes more pronounced. This dynamic, combined with \textit{reoxygenation}, which reactivates the highly proliferative behavior of cells in the low epigenetic band, leads to a slight decrease in the average phenotypic trait. However, this change is not stable, as we can infer from previous analyses, and there are two potential drawbacks. 
The first concern is that, initially, \textit{reoxygenation} and the availability of space, facilitated by the high kill rate, create conditions for rapid tumour regrowth. The second concern arises from the rapid overpopulation, which causes the oxygen profile to rapidly decline and, as a result, selection of resistant cells, and formation of necrotic areas.} 
In the same veins, the dynamics observed in the case of low oxygenation are even more interesting in terms of tailored radiotherapy protocols. 
As we can observe, the impact of the treatment's selective pressure is less evident. In fact, we notice variations of lower intensity compared to the previous scenario as the dosage increases in terms of both radiosensitivity and epigenetic firmness  {(see panels (C-FV) and (C-HV) in Fig. \ref{fig:comp}). The only noticeable effect is seen on radiosensitivity for a dose of $74$ Gy, which can be primarily attributed to the intensity of \textit{reoxygenation}. This particular case is evidently more challenging in terms of treatment effectiveness. 
The dose of $46$ Gy has almost no impact on cell killing, and the dose of $60$ Gy proves to be ineffective as well. The only dose that can effectively eliminate a substantial number of cells is $72$ Gy, but it also carries potential risks. Even in the presence of reoxygenation, the optimal epigenetic signature tends to be the high one across most of the domain (as we saw in Fig. \ref{fig:simu1})} . This supports the idea to target hypoxic areas with higher doses since no stronger selection with respect treatment resistance is observed,  {however, we pay special attention to the possibility of critical and highly resistant relapse in cases where complete eradication is not achieved. From this perspective, we hypothesize that the model's sensitivity in capturing these dynamics lays the groundwork for investigating various therapeutic protocols using this modeling approach. For instance, we can explore scenarios involving innovative techniques such as Stereotactic Body Radiation Therapy (SBRT) which involves delivering a small number of high radiation doses to a specific target volume using highly accurate equipment with the goal to optimize cancer control while minimizing adverse effects on healthy tissues. see e.g. \cite{lo2010stereotactic}.}

\subsection{ {Heterogeneous tissue oxygenation}}
\label{subsec:Heterogeneous vasculature}
\begin{figure}
\includegraphics[width=1\linewidth]{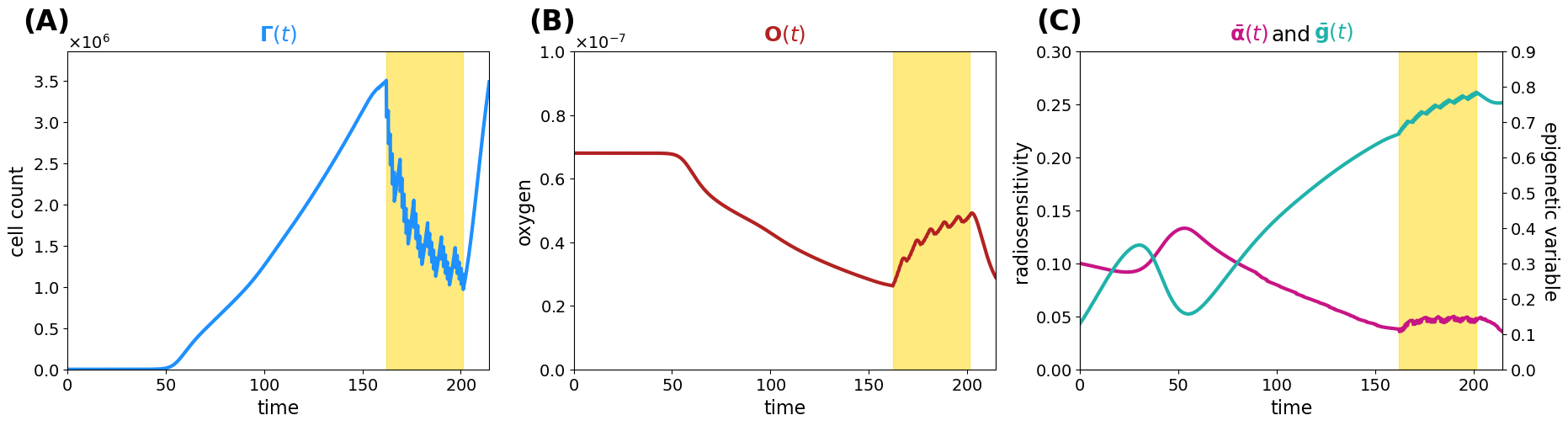}
\includegraphics[width=1\linewidth]{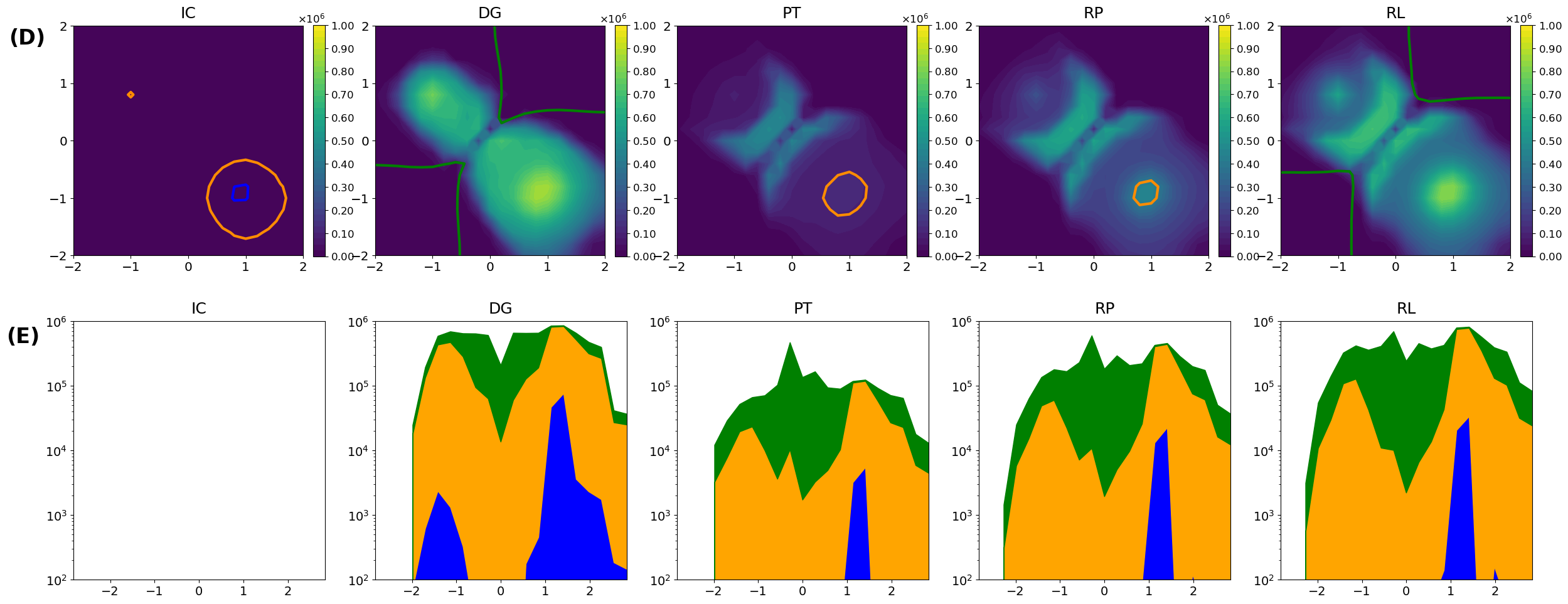}
\caption{Simulation of tumour mass, growing from the center of the domain, with oxygen provided by the three  {source} layout $\Upsilon^{3V}$. 
Plots description as in caption of Fig. \ref{fig:simu1}.
}
\label{fig:fagiolo}
\end{figure}

In this last experiment, 
we focus on a  {heterogeneous tissue oxygenation} considering 
three  {oxygen sources} in the $\Upsilon^{3V}$ configuration.
With such a layout, the most oxygenated areas are concentrated at the ends of the domain antibisector, leaving a condition of low oxygenation between them and along the bisector (see plot (D-IC) of Fig. \ref{fig:fagiolo} where the optimal oxygenation areas are outlined).
Thus, keeping the starting point of the tumour unchanged at the center of the domain, the optimal trait, determined by the initial condition of the oxygen, is in the high epigenetic band.
 {Consequently, the highly proliferative epigenetic characteristic that defines the initial tumour mass is suboptimal in terms of selection. This becomes evident when analyzing plot (C) in Fig. \ref{fig:fagiolo} in comparison to the cell count evolution shown in plot (A). Initially, the average epigenetic trait is very low, but it gradually increases during the first phase due to selective pressure. During this period, the cell count experiences slow growth. Subsequently, cancer cells reach more oxygenated areas where lower epigenetic traits are favored. Around day $50$, the average epigenetic trait reaches a local minimum, and the cell count starts to rapidly increase. This leads to a significant oxygen consumption until the moment of diagnosis. 
By comparing plots (D-IC) and (D-DG) in Fig. \ref{fig:fagiolo}, we observe that at the initial time, there are optimal areas for all epigenetic traits and no necrotic regions. However, at the time of diagnosis, the oxygen has been consumed by cancer cells (consistently with plot (B)), resulting in a wide area of optimality only for high epigenetic traits, accompanied by some necrotic areas around the corner of the bisector.
}
Then, when therapy is applied, despite an initial high mortality rate due to radiotherapy, its effectiveness decreases significantly in the second phase of administration, see plot (A) of Fig. \ref{fig:fagiolo}.
The low level of oxygenation in the central area of the domain maintains an important effect both of direct reduction of the therapeutic efficacy in this area, and of selection of treatment-resistant epigenetic traits, plot (D)-PR of Fig. \ref{fig:fagiolo}.
Thus, therapy speeds up and assists the same dynamic that occurs due to environmental selection. \\
 {In the areas close to the oxygen sources, cells exhibit low epigenetic traits, leading to higher mortality rates. The significant death of these cells (as depicted in plot (D-PT) where two big holes appear corresponding to the higher density zone in (D-DG)) triggers tissue \textit{reoxygenation}, coherently with the oxygen profile during therapy in plot (B). This phenomenon is illustrated in plot (C) by the observation that, despite the increase in average epigenetic traits, the radiosensitivity fluctuates around a constant value during therapy administration, counterbalanced by the rise in oxygen levels.}

Analyzing the temporal phases of each epigenetic band,  
 plots (D) and (E) of Fig. \ref{fig:fagiolo}, 
it is clear how the overlap of selective dynamics and therapy can modify the conformation of the tumour mass. In fact, comparing the second and last image,  
which refer to the moment of diagnosis (DG column) and to the moment of relapse (RL column), while 
the tumour has the same size in terms of numerosity, it shows many differences both from the point of view of the shape of the tumour and of the composition.
Indeed, in a first phase (until diagnosis time), high epigenetic band has the selective fitness advantage (except from the small areas contoured in orange and blue), but it suffers from the proliferative dominance of lower epigenetic signatures.
During therapy, the overlap of therapeutic resistance and environmental selection 
gives high epigenetic band an advantage, which is largely maintained in the \textit{repopulation} phase, with the only exception of the lower right corner, where the fitness of low epigenetic band promotes an accumulation of highly proliferative cells, plot (E)-RP of Fig. \ref{fig:fagiolo}.
When the relapse occurs,  from an epigenetic composition point of view, low epigenetic band cells kept a peak nearby the two oxygen sources close together, but are almost absent at the center of the domain and near the other source, differently from the previous times.

The above results highlight how a heterogeneous vasculature may lead to profound differences between the epigenetic composition of the tumour and its geometric characterization at the time of diagnosis and relapse. In conclusion, this experiment therefore shows the deep impact of therapy   
on the environment and on the characteristics of the tumour and it highlights, even more than the previous experiments, the potential of the model as a basis for therapeutic optimization strategies based on knowledge and predictive ability of the development of the mass.

\section{Conclusions}
\label{sec:Conclusions}
In this work, we presented a mathematical approach to explore how low oxygen levels and hypoxia-associated tumour cell adaptions affect radiotherapy efficiency in the specific case of solid tumours. Specifically, we compared the effect of tumour microenvironment in the case of an efficient or inefficient tumour vasculature evaluating (\textit{i}) how it can influence the heterogeneity in terms of proliferative potential of tumour cells and (\textit{ii}) how its evolution could strongly influence the treatment success. The rationale of the work was (\textit{i}) to identify the tumour regions composed by cells with low proliferative rate that are intrinsically more resistant to radiotherapy action, (\textit{ii}) to study the consequences of the treatment in influencing their geometrical characterization and (\textit{iii}) to investigate if and how they can be potentially separately treated to maximize the tumour response, toward a tailoring of radiotherapy protocols and \textit{dose painting} perspective, 
\cite{grimes2017hypoxia}.\\
The results 
show how the proposed approach is, first of all, able to reproduce the biological effect of irradiation as the result of both the total dose delivered and the physiological conditions in which it is applied. Moreover, the findings support the ability of the model to mirror specific eco-evolutionary features of different tumour masses, making predictions based on conditions that can widely range between patients.\\
Specifically, our tool suggests how three of the $6R$ that characterize tumour response after radiotherapy administration, \textit{repopulation}, \textit{reoxygenation} and \textit{radiosensitivity}, could display 
different dynamics in dependence on tumour oxygenation and the consequent distinct tumour-host interaction, \cite{rakotomalala2021hypoxia}. Summarizing, two relevant dynamics from a clinical point of view are kept by the model. Firstly, coherently with the clinically observed inter-patient variability in terms of therapy response, the nominal tumour size alone is insufficient to predict growth dynamics and that the definition of personalized indexes is needed to define an efficient therapeutic plan. In this respect, the model results suggest that two patients that present a similar tumour volume could have a distinct tumour–host co-evolution history, which results in different responses to the same radiotherapy protocol, \cite{poleszczuk2018predicting}. In this respect, in the two cases analysed, it is significantly distinct the in-time evolution of the radiosensitivity of the mass, guided by both the different radio-induced reoxygenation and epigenetic composition. Our predictions show indeed that, in the case of high oxygenation, \textit{reoxygenation} is sufficient to prompt the mass towards an higher sensibility to the treatment; at the contrary, in the case of low oxygenation, a degenerating dynamics towards a more resistant mass can be observed, even in presence of higher quantity of available oxygen, highlighting the central role of epigenetic heterogeneity in tumour therapy response. Secondly, to maximize the effect of the treatment in terms of a balance between the portion killed and the selective bottleneck induced, the choice of the dose amount administrated turns out to be necessary related to tumour oxygenation. The model results suggest indeed that, under oxygenation conditions that do not strongly affect the effectiveness of therapy, the heterogeneity of the mass plays a crucial role in the development of treatment resistance; at the highest radiation dosage, a marked epigenetic shift towards resistant epigenetic traits and a rapid decrease in the radiosensitivity of the mass are indeed observe. At the contrary, in strongly hypoxic conditions, the predictions reveal a greater reduction in mass for higher doses but, at the same time, not an equally large variation in terms of composition and radiosensitivity compared to those observed at lower doses, supporting the idea of tailoring radiotherapy protocols and \textit{dose painting} that consists in delivery selective boosting dose to radio-resistant (in this specific case hypoxic) regions.\\
Supported by the results, we speculate that the sensibility of the model in catching these dynamics potentially paves the way to investigate, via this modelling approach, different administration scenarios in the case
of, for example, innovative techniques as Stereotactic Body Radiation Therapy (SBRT) in which a small
number of high doses of radiation are delivered to a target volume using highly accurate equipment in
order to maximize cancer control, \cite{grimes2017hypoxia}. As natural evolution, future studies will focus on the model outcomes varying the total dosage, the target regions but additionally the fractionation scheme. There is indeed biological evidence that alternative radiation fractionation protocols sometimes improve the outcome while worsen in others cases; altered schemes, such as hyperfractionation, accelerated fractionation and hypofractionation, have	been suggested as alternatives for certain indications (\cite{kaaver1999stochastic,tucker1989effect}).\\
 This, in addition to the already shown results, potentially allows to exploit our tool to investigate possible therapeutic strategies to optimize the radiotherapy outcome in light of the epigenetic and geometric inhomogeneities, considering the inter-patients variability experimentally observed.

 {Finally, an intriguing perspective arises when interpreting the results of our paper in relation to the concept of "cancer oncospace", as introduced in \cite{aguade2023oncospace}. This concept encompasses a multidimensional space that captures the intricate and shared characteristics of cancer heterogeneity. Specifically, the oncospace framework aims to investigate the underlying mechanisms contributing to tumour development and progression, encompassing ecological, evolutionary, and developmental factors. \\
%
Our paper's findings provide valuable insights into the dynamics of tumour response to radiotherapy, taking into account factors such as tumour oxygenation and epigenetic composition. They reveal distinct co-evolution histories and responses to the same treatment protocol among different tumour masses. These findings underscore the significance of personalized indices and the consideration of inter-patient variability when designing efficient therapeutic plans. Overall, these results align with the principles of cancer oncospace by emphasizing the importance of comprehending the multidimensional nature of cancer, including factors like tumour microenvironment, heterogeneity, and personalized treatment approaches, to optimize therapy outcomes. 
}

\section*{Conflict of Interest Statement}

The authors declare that the research was conducted in the absence of any commercial or financial relationships that could be construed as a potential conflict of interest.

\section*{Author Contributions}
\noindent\textbf{G. C.} : conceptualization, methodology, software, writing, visualization;\\
\textbf{G. F.} : conceptualization, methodology, writing, visualization;\\
\textbf{M. E. D.} : conceptualization, methodology, writing, visualization, supervision.

\section*{Funding}
This research was partially supported by the Italian Ministry of Education, University and Research (MIUR) through the ``Dipartimenti di Eccellenza'' Programme (2018-2022) -- Dipartimento di Scienze Matematiche ``G. L. Lagrange'', Politecnico di Torino (CUP: E11G18000350001). 

\section*{Acknowledgments}
 All authors are members of GNFM (Gruppo Nazionale per la Fisica Matematica) of INdAM (Istituto Nazionale di Alta Matematica).



\bibliographystyle{ieeetr}
\bibliography{bibliography}
\end{document}